\documentclass[%
 reprint,
superscriptaddress,
nofootinbib,
 amsmath,amssymb,
 aps, preprintnumbers,
floatfix
]{revtex4-2}

\usepackage{graphicx}
\usepackage{dcolumn}
\usepackage{bm}
\usepackage{hyperref}
\usepackage[modulo]{lineno}
\usepackage[caption=false]{subfig}
\usepackage{multirow}
\usepackage{diagbox}
\usepackage{tabularx}
\usepackage[dvipsnames]{xcolor}
\usepackage{booktabs}
\usepackage{floatrow}
\usepackage{comment}

\usepackage{graphicx}
\usepackage[rightcaption]{sidecap}
\sidecaptionvpos{figure}{c}

\newfloatcommand{capbtabbox}{table}[][\FBwidth]

\usepackage{enumitem}

\usepackage{xspace}
\usepackage{soul}

\usepackage{lineno}

\RequirePackage[normalem]{ulem} 
\RequirePackage{color}\definecolor{RED}{rgb}{1,0,0}\definecolor{BLUE}{rgb}{0,0,1} 

\newcommand{\unit}[1]{\ensuremath{\mathrm{\,#1}}\xspace}

\newcommand{\e}{\unit{e^{-}}}

\newcolumntype{Y}{>{\centering\arraybackslash}X}

\begin{document}

\preprint{FERMILAB-PUB-26-0349-PPD} 

\modulolinenumbers[1]

\title{Characterization of Spurious Charge in SENSEI Skipper-CCDs
}

\author{Yikai Wu}
\affiliation{\normalsize\it 
C.N.~Yang Institute for Theoretical Physics, Stony Brook University, Stony Brook, NY 11794, USA}
\affiliation{\normalsize\it 
Department of Physics and Astronomy, Stony Brook University, Stony Brook, NY 11794, USA}

\author{Ansh Desai}
\affiliation{\normalsize\it 
Department of Physics and Institute for Fundamental Science, University of Oregon, Eugene, Oregon 97403, USA}

\author{Sho Uemura}
\affiliation{\normalsize\it 
Fermi National Accelerator Laboratory, PO Box 500, Batavia IL, 60510, USA}

\author{Ana M. Botti}
\affiliation{\normalsize\it
Département de Physique, Université de Montréal, Montréal, Québec H3C 3J7, Canada
}

\author{Brenda A. Cervantes-Vergara}
\affiliation{\normalsize\it 
Fermi National Accelerator Laboratory, PO Box 500, Batavia IL, 60510, USA}

\author{Fernando Chierchie}
\affiliation{\normalsize\it 
Instituto de Inv. en Ing. Eléctrica (IIIE), Dpto. de Ingeniería Eléctrica y de Computadoras (DIEC), Universidad Nacional del Sur (UNS)-CONICET, Bahía Blanca, Argentina
}

\author{Alex Drlica-Wagner}
\affiliation{\normalsize\it 
Fermi National Accelerator Laboratory, PO Box 500, Batavia IL, 60510, USA}
\affiliation{\normalsize\it  Department of Astronomy and Astrophysics, University of Chicago, Chicago IL 60637, USA}

\author{Rouven Essig}
\affiliation{\normalsize\it 
C.N.~Yang Institute for Theoretical Physics, Stony Brook University, Stony Brook, NY 11794, USA}

\author{Juan Estrada}
\affiliation{\normalsize\it 
Brookhaven National Laboratory, P.O. Box 5000 Upton, NY 11973-5000}
\affiliation{\normalsize\it 
Fermi National Accelerator Laboratory, PO Box 500, Batavia IL, 60510, USA}
\affiliation{\normalsize\it  Department of Astronomy and Astrophysics, University of Chicago, Chicago IL 60637, USA}

\author{Erez Etzion}
\affiliation{\normalsize\it 
 School of Physics and Astronomy, 
 Tel-Aviv University, Tel-Aviv 69978, Israel}

\author{Guillermo Fernandez Moroni}
\affiliation{\normalsize\it 
Fermi National Accelerator Laboratory, PO Box 500, Batavia IL, 60510, USA}
\affiliation{\normalsize\it  Department of Astronomy and Astrophysics, University of Chicago, Chicago IL 60637, USA}

\author{Miqueas Gamero}
\affiliation{\normalsize\it 
Instituto de Inv. en Ing. Eléctrica (IIIE), Dpto. de Ingeniería Eléctrica y de Computadoras (DIEC), Universidad Nacional del Sur (UNS)-CONICET, Bahía Blanca, Argentina
}

\author{Stephen E. Holland}
\affiliation{\normalsize\it 
Lawrence Berkeley National Laboratory, One Cyclotron Road, Berkeley, California 94720, USA}

\author{Ian Lawson}
\affiliation{\normalsize\it SNOLAB, Lively, ON P3Y 1N2, Canada}
\affiliation{\normalsize\it School of Natural Sciences, Laurentian University, Sudbury, P3E 2C6, Canada}

\author{Steffon Luoma}
\affiliation{\normalsize\it SNOLAB, Lively, ON P3Y 1N2, Canada}

\author{Nathan A. Saffold}
\affiliation{\normalsize\it 
Fermi National Accelerator Laboratory, PO Box 500, Batavia IL, 60510, USA}

\author{Miguel Sofo-Haro}
\affiliation{\normalsize\it 
Instituto de Física E. Gaviola (IFEG), CONICET, Ciudad Universitaria, Córdoba 5000, Argentina
}
\affiliation{\normalsize\it 
Reactor Nuclear RA0 (CNEA), Universidad Nacional de Córdoba, Córdoba, Argentina.
}

\author{Javier Tiffenberg}
\affiliation{\normalsize\it 
Fermi National Accelerator Laboratory, PO Box 500, Batavia IL, 60510, USA}

\author{Tomer Volansky}
\affiliation{\normalsize\it 
School of Physics and Astronomy, 
 Tel-Aviv University, Tel-Aviv 69978, Israel}

\date{\today}

\begin{abstract}
Skipper Charge-Coupled Devices (Skipper-CCDs) are a leading technology in the search for sub-GeV dark matter and coherent elastic neutrino-nucleus scattering. A key background for rare-event searches with these detectors arises from ``spurious charge''---single-electron events generated when charges are transferred through the active region to the serial register, and across the serial register to the readout stage. We present a characterization of spurious charge in both the active region and the serial register of SENSEI Skipper-CCDs, and show that, in a well-shielded low-background environment, the dominant contribution originates in the serial register during Skipper readout, when horizontal clocks are held at constant voltage between pixel transfers. 
Motivated by this finding, we develop a ``tri-level'' clocking scheme in which the held-low phase is raised to an intermediate voltage during readout to suppress trap-mediated charge generation.
Using the SENSEI detector near the MINOS cavern, we measure a serial-register single-electron density of $(2.9\pm0.1)\times10^{-5}\unit{\e/pixel/image}$ under standard SENSEI readout conditions, reduced to $(4.0\pm 0.4)\times10^{-6}\unit{\e/pixel/image}$ with tri-level clocking---a factor of $\sim$7 improvement. This technique offers a promising path to lower backgrounds in current and future Skipper-CCD experiments.
\end{abstract}

\maketitle

\section{\label{sec:intro}Introduction}
Skipper Charge-Coupled Devices (Skipper-CCDs), leveraging single-charge resolution, low background, and high spatial resolution, have been one of the leading technologies for dark matter and coherent neutrino scattering~\cite{skipper1e,sensei_2020,sensei_snolab,sensei_modulation,damic_modu_2023,damic_2024,CONNIE_2024}. This technology has enabled the SENSEI and DAMIC-M collaborations to set world-leading constraints on sub-GeV dark matter~\cite{sensei_1e_rate,damic_probing,damic_new_modulation}.
However, ongoing searches are limited by low-energy background events, the majority of which arise from single-electron events and 
their pileup.
Single-electron events can be categorized into two groups: (1) exposure-dependent and (2) exposure-independent events.
Several efforts have been made to reduce exposure-dependent single-electron events, which are mainly produced by environmental radiation, blackbody radiation of warm materials surrounding the Skipper-CCDs, and dark current~\cite{sensei_2020,Du:2020ldo,lee_ccd,sensei_mariano,sensei_1e_rate}.
The lowest single-electron rate obtained is $(1.39\pm0.11)\times10^{-5}\unit{\e/pixel/day}$, while the exposure-independent single-electron density is at least $\sim5\times10^{-5}\unit{\e/pixel/image}$~\cite{sensei_1e_rate}.
A significant exposure-independent source of single-electron events is spurious charge induced by clocking produced during Skipper-CCD readout operations, also referred as clock-induced charge~\cite{sensei_mariano,lee_ccd}. 
Therefore, it is important to understand and mitigate spurious charge in order to further improve sensitivity of low background experiments using Skipper-CCDs.

In this paper, we present a characterization of spurious charge produced by both vertical transfers in the active region and horizontal transfers in the serial register of SENSEI Skipper-CCDs. The paper is organized as follows. Section~\ref{sec:sc} describes a model of spurious charge generation in Skipper-CCDs. Section~\ref{sec:setup} outlines the SENSEI Skipper-CCD format and experimental setup for measuring spurious charge. Section~\ref{sec:meas} details the Skipper-CCD initialization procedure and the methods to measure the spurious charge both in the active region (``active region spurious charge,'' abbreviated ARSC) and in the serial register (``serial register spurious charge,'' abbreviated SRSC). Section~\ref{sec:arsc} presents the characterization of the ARSC and the impact of Skipper-CCD initialization. Section~\ref{sec:srsc} presents the characterization of SRSC as well as the result of a modified tri-level clocking scheme for reducing serial register spurious charge in Skipper-CCDs.

\section{\label{sec:sc} Spurious charge in Skipper-CCDs}

Skipper-CCDs function as pixelated calorimeters that measure electron-hole pairs generated by ionizing particles interacting with the bulk of the device.
A substrate voltage applied at the backside of the Skipper-CCD is sufficient to fully deplete it, separating holes and electrons throughout the bulk; for the ``p-channel'' CCDs used by SENSEI, the holes, which serve as the signal charge (denoted as \e, despite not being electrons), drift toward the front surface, while the electrons drift toward the backside of the Skipper-CCD and are drained through the substrate bias contact.
The holes are collected in a potential minimum, the ``buried channel,'' approximately $1\unit{\mu m}$ beneath the $\unit{SiO_2}$ insulator that forms the front surface; the buried channel protects the holes from charge traps that may exist at the $\unit{Si-SiO_2}$ interface.

During readout, holes collected in the pixel grid of the active region are transferred row-by-row to a serial register at the edge of the sensor and then transferred column-by-column to a readout stage. 
Each pixel in the active region or serial register is covered by three overlapping gates, called ``clocks'' or ``phases.''
When the phases are clocked between high and low voltage states in sequence, they produce a moving potential well that transfers charges to the next pixel in the active region or the serial register.
The readout stage consists of a floating-gate amplifier, where repeated non-destructive measurements of the number of holes in each charge packet are performed to reach sub-electron noise~\cite{skipper1e}; this procedure is referred to as ``Skipper readout.''
Each Skipper-CCD has two serial registers located at opposing edges of the Skipper-CCD and four identical readout stages located at the Skipper-CCD corners. The four readout stages allow parallel readout of four distinct quadrants.

\begin{figure}[!t]
    \centering
    \includegraphics[width=1\textwidth]{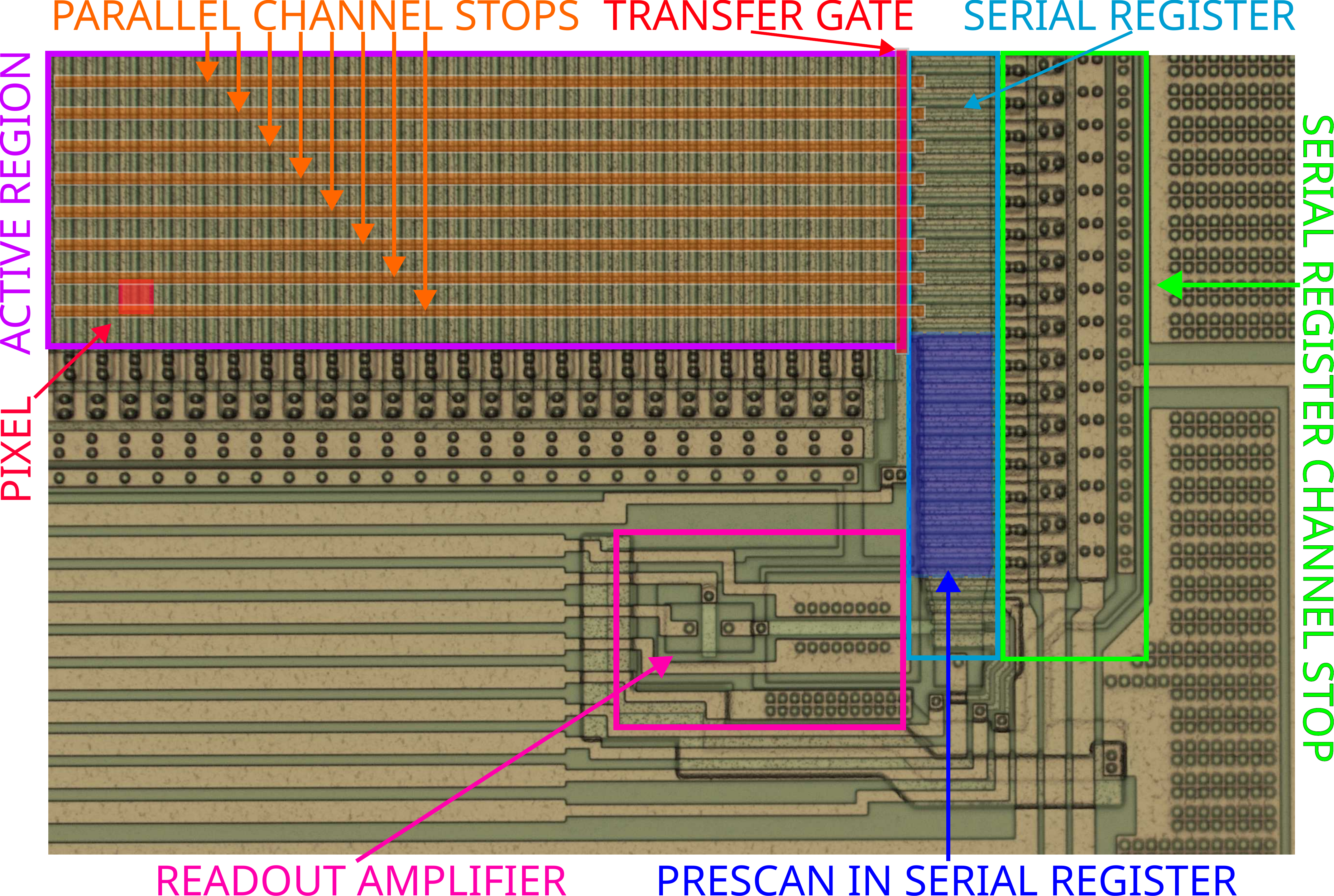}
    \caption{Microscope image showing one corner of a Skipper-CCD, illustrating the layout of the active region, serial register, and readout amplifier. An eight-pixel extension of the serial register beyond the active region constitutes the prescan region. In this region, the serial register lacks adjacent parallel channel stops and is isolated from the active region columns.}
    \label{fig:ccd_structure}
\end{figure}

Fig.~\ref{fig:ccd_structure} shows a microscope image of a corner of the Skipper-CCD.
In addition to the pixels and serial register that collect and transfer holes, the adjacent n-type implant ``channel stops'' are highlighted.
In the active region (and the transfer gate separating it from the serial register), pixel columns are separated by narrow n-type implants (denoted parallel channel stops) to prevent holes collected in the buried channels from diffusing into adjacent columns; these channel stops are electrically isolated and end at the serial register. A wide n-type implant (denoted serial register channel stop) is present near each serial register underneath the metal lines for controlling horizontal clock voltages; this channel stop is connected to a fixed bias~\cite{steve_2009}.
Unlike most CCDs where it is possible to ``invert'' some or all of the CCD surface during operation to flood the $\unit{Si-SiO_2}$ interface with electrons from the channel stops~\cite{ccd_pinning}, any electrons at the surface of these fully depleted CCDs are largely confined to the channel stops except during CCD initialization, and the channel stops themselves can be purged of electrons~\cite{steve_2006}.

\begin{figure}[!t]
    \centering
    \includegraphics[scale=0.51]{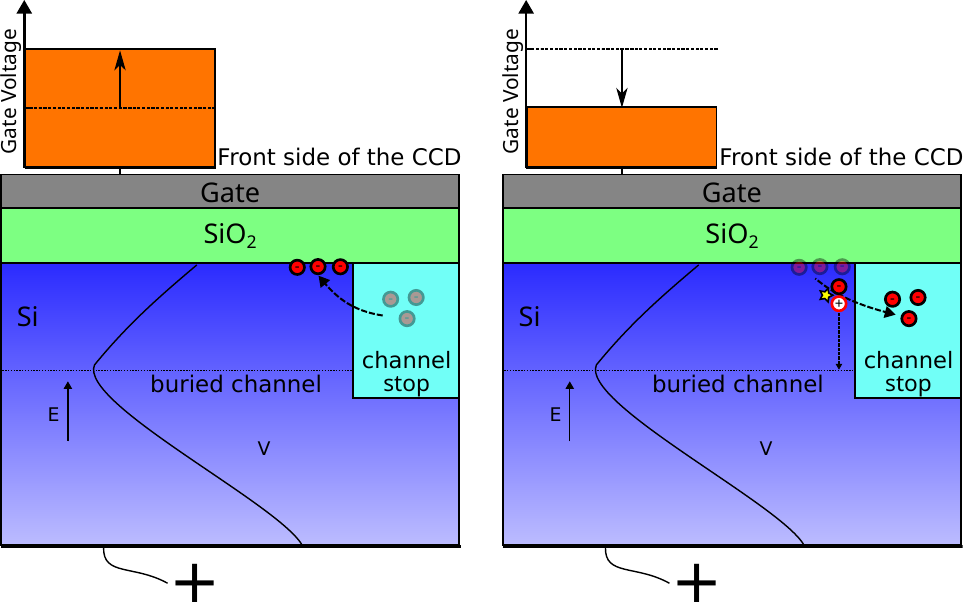}
    \caption{Schematic diagram of spurious charge generation in p-channel Skipper-CCDs. \textbf{Left}: During the high gate voltage state, electrons from channel stops are attracted to the \unit{Si-SiO_2} interface, where some are captured by defects. \textbf{Right}: During and after the subsequent gate transition to the low voltage state, trapped electrons that are released are accelerated back to the channel stops by the large potential gradient between interface and channel stops. With enough energy, these electrons induce impact ionization resulting in the generation of holes. These holes are collected in the buried channel as spurious charge.}
    \label{fig:sc_diagram}
\end{figure}

The hypothesized mechanism of spurious charge generation in a p-channel Skipper-CCD is shown schematically in Fig.~\ref{fig:sc_diagram}.
Similar to the mechanism for CCDs that operate in full or partial inversion~\cite{janesick2001scientific}, it depends on electrons that become trapped at a phase's $\unit{Si-SiO_2}$ interface when the gate voltage is high and detrap during or after the subsequent lowering of the gate voltage.
In contrast to invertible CCDs, where spurious charge appears when the gate voltage exceeds the threshold to invert the entire phase~\cite{janesick2001scientific,emccd}, we speculate that the spurious charge we observe is generated at the fringes of the channel stops.
When the gate voltage of a pixel (either vertical or horizontal) is raised to its high state, more of the electrons in the channel stop are pulled towards that phase.
This results in some region of the $\unit{Si-SiO_2}$ interface near the edge of the channel stop being newly exposed to electrons. 
A fraction of these electrons are captured by defects at the $\unit{Si-SiO_2}$ interface.

When the gate voltage is lowered, electrons are repelled back to channel stops.
Free electrons move during the voltage change and see low electric fields, and do not produce significant spurious charge.
However, trapped electrons may be released at later times after the gate voltage is fully lowered, and the high potential gradient can be sufficient to induce impact ionization.
The holes from impact ionization are then collected in the buried channel and are indistinguishable from ionization events produced in the bulk silicon of the Skipper-CCD.

By this model, we should expect to see that the spurious charge rate declines steadily after the gate voltage goes low (as traps decay), but that this process repeats after the gate voltage returns to its high value (refilling the traps).
Furthermore, at an operating temperature of $\sim135\unit{K}$, the emission time constant is several orders of magnitude larger than the capture time constant~\cite{SRH}.
Consequently, the trap occupancy (and thus spurious charge) is not sensitive to the duration for which the gate voltage is high.

Based on this model, spurious charge generation requires a neighboring channel stop with electrons, and the amount of spurious charge scales with the number of pixel transfers $N_{\text{transfer}}$ and increases with the peak-to-peak voltage $\Delta V$, denoted as ``clock swing.''
The total amount of spurious charge generated in time $t$ after a gate voltage is lowered $R_{\text{SC}}(t)$ is then given by:
\begin{equation}
    R_{\text{SC}}(t) \propto  \int_{0}^{\infty} f(\tau_{\text{e}})(1-e^{-\frac{t}{\tau_{\text{e}}}}) \, d\tau_{\text{e}}\ ,
    \label{eq:sc_model}
\end{equation}
where $\tau_{\text{e}}$ is the emission time constant, and $f(\tau_{\text{e}})$ is the distribution of trap emission time constants.

\section{\label{sec:setup}Experimental Setup}

The CCDs used in this work to characterize spurious charge are identical to the Skipper-CCDs used to obtain previous results with SENSEI detectors operating at SNOLAB and near MINOS~\cite{sensei_snolab,sensei_modulation,sensei_1e_rate}.
They are three-phase, fully-depleted p-channel Skipper-CCDs designed by Lawrence Berkeley National Laboratory and fabricated on high resistivity $n$-type silicon at Teledyne/DALSA.
These Skipper-CCDs have a thickness of $665\unit{\mu m}$ and feature pixels with a size of $15\times15\unit{\mu m^2}$.
Two types of Skipper-CCDs are used which differ only in the dimensions of the pixel array.
The ``C-module,'' used in \cite{sensei_snolab,sensei_modulation,sensei_1e_rate}, has a pixel array of $6144\times1024$ pixels where the serial registers are located along the long edges of the sensor; each serial register of a quadrant has $3080$ pixels including an eight-pixel prescan region. 
The ``skinny module,'' used in \cite{sensei_2020,sensei_mariano}, has a pixel array of $886\times6144$ pixels, and the serial registers are located along the short edges of the sensor; each serial register of a quadrant has $451$ pixels including an eight-pixel prescan region.

For all measurements in this work, the Skipper-CCD is housed in a copper tray sealed with copper tape to minimize single-electron events from infrared photons emitted by warm material surrounding the tray~\cite{sensei_1e_rate}. A single low-threshold acquisition board~\cite{lta} is used to read out each Skipper-CCD.
A substrate bias of $70\unit{V}$ is used to fully deplete the bulk silicon in the Skipper-CCD, and the system is cooled to $135\unit{K}$ to minimize dark current.
For normal SENSEI operations, both vertical and horizontal clock swings are $2.25\unit{V}$.

Three detector configurations are used for characterizing spurious charge in this paper:
\begin{enumerate}
    \item Underground setup with C-module: This setup is identical to the SENSEI@MINOS detector used in previous single-electron rate measurements~\cite{sensei_1e_rate}, although of the two installed C-module Skipper-CCDs, only one is functional.
This detector is located $\sim{100\unit{m}}$ underground in the MINOS cavern at Fermilab and surrounded by lead shielding to reduce cosmic rays and other high-energy events.
The readout electronics contain RC filters with the standard time constants used in SENSEI: $10\unit{\mu s}$ for vertical clocks and $4.7\unit{\mu s}$ for horizontal clocks.
These filters are intended to shape the clock waveforms to reduce spurious charge~\cite{janesick2001scientific}.
    \item Underground setup with skinny modules: The SENSEI@MINOS apparatus has recently been modified to house four skinny-module Skipper-CCDs. All other details remain unchanged.

\item Surface setup with C-module: A single C-module Skipper-CCD is housed in a stainless steel dewar at a Fermilab surface facility.
This setup lacks lead shielding, but the readout electronics allow the time constants of the RC filters to be modified easily.
\end{enumerate} 

While background radiation does not affect the mechanism of spurious charge generation in Skipper-CCDs, the low-background environment of the SENSEI@MINOS detector enables effective measurement of the spurious charge at the levels relevant for rare-event searches, typically on the order of $10^{-5}\unit{\e/pixel/image}$.

\section{\label{sec:meas}Spurious charge measurement protocol}
To characterize spurious charge in Skipper-CCDs, we utilize a pumping technique where active region spurious charge (ARSC) or serial register spurious charge (SRSC) is generated by repeated clocking. 
While the pumping mechanism is similar for both regions, the readout and data processing differ. The details for each region are described in the respective subsections below.

\subsection{Initialization of Skipper-CCDs}
Before each measurement, the Skipper-CCD is initialized following standard SENSEI procedures to ensure a low baseline of dark counts. The initialization procedure is as follows:

\begin{enumerate}
    \item Temperature cycle: This is performed after each power cycle of the readout electronics. In this procedure, the ``erase'' and ``e-purge'' procedures (described next) are run repeatedly during a ramp down from an elevated temperature of $220\unit{K}$ to the nominal temperature of $135\unit{K}$. This procedure minimizes dark current.
    \item Erase and e-purge: This is performed during the temperature cycle and additionally as needed to re-initialize the dark current to a baseline state.
    In the erase procedure, the substrate voltage is ramped down from $70\unit{V}$ to $0\unit{V}$ and a set of clock voltages are increased to $9\unit{V}$, which floods the surface of the Skipper-CCD with electrons. This state is maintained for $\ge2\unit{s}$. The substrate voltage is then ramped back to $70\unit{V}$; the clock voltages are restored to their normal operating values when the substrate voltage reaches $10\unit{V}$.
    The erase fills traps that would otherwise generate dark current~\cite{steve_2003,burkegajar}.
    Immediately following the erase, an e-purge procedure is performed, where a set of clock voltages are lowered to $-9\unit{V}$ for $\ge2\unit{s}$ while maintaining the substrate voltage at $70\unit{V}$. The e-purge is intended to purge the populations of electrons in the parallel channel stops~\cite{LBNL_CCD_Manual,e2v_purge}.
    Both the erase and e-purge can be applied with various sets of clock voltages. In SENSEI, both the erase and e-purge procedures are usually done with all clock voltages being manipulated. 
    In this work, we consider in addition to this standard e-purge procedure a set of different e-purge procedures, in which different set of clocks are manipulated during e-purge.  In particular, we define the following two fundamental cases:
\begin{itemize}
\item ``all-clocks purge:'' where all clocks are lowered as in the standard SENSEI procedure.
\item ``vertical-only purge:'' only the vertical clocks are lowered and all other clock voltages remain at their nominal values~\cite{LBNL_CCD_Manual}.
\end{itemize}
We will see that a Skipper-CCD initialized with the vertical-only purge procedure can generate drastically lower amounts of spurious charge in both the active region and serial register when compared to a Skipper-CCD initialized with the all-clocks purge procedure. 
\item Cleaning: Before taking each image, the charges in the active region and serial register are shifted to readout amplifiers and drained repeatedly to remove accumulated charges from background radiation, as well as to remove single electrons from charge traps with long emission time constants. This ensures a clean image before the spurious charge measurement.
\end{enumerate}

\subsection{Pumping procedures}
The pumping procedures involve generating a controlled amount of ARSC or SRSC by clocking the vertical or horizontal gate voltages respectively. The clocking is performed with either pixel-pumping or phase-pumping: 
\begin{enumerate}
    \item[(1)] In pixel-pumping, holes are transferred back and forth between adjacent pixels. Each cycle involves two transfers to move the charge packet to the center of a neighboring pixel and then back to the original position.
    \item[(2)] The phase-pumping involves toggling only a single clock phase while the other two phases are held constant to avoid charge transfer. Each phase-transition cycle is counted as one pump.
\end{enumerate}
To isolate the spurious charge generated during pumping from other single-electron backgrounds, such as dark current generated during exposure and spurious charge generated during readout, we take multiple measurements while varying only the number of transfers or pumps ($n$).
The total amount of charge in the active region or the serial register follows a Poisson distribution with a mean $\mu$ (in units of $\unit{\e/pixel/image}$) that scales linearly with the number of transfers or pumps:
\begin{equation}
    \mu(n)=R_{\text{SC}}N_{\text{bin}}n+\mu_{\text{other}}\, ,
    \label{eq:sc_fit}
\end{equation}
where $R_{\text{SC}}$ is the spurious charge rate (in unit of $\unit{\e/pixel/transfer}$ or $\unit{\e/pixel/pump}$), $\mu_{\text{other}}$ characterizes single-electron backgrounds from dark current, amplifier light, and spurious charge generated during readout, and $N_{\text{bin}}$ is the binning factor when multiple pixels are combined into a ``superpixel'' before charge measurement (when we refer below to $A\times B$ binning, it means combining $A$ columns and $B$ rows into a superpixel, with a resulting binning factor of $N_{\text{bin}}=AB$). The spurious charge rate $R_{\text{SC}}$ is extracted by performing a linear fit of $\mu$ versus $n$.

\subsection{Readout and Data selection}
\subsubsection{Active region spurious charge}
For the ARSC measurement, the active region along with additional overscan regions are read out through all four amplifiers following pumping in the active region.
The standard SENSEI calibration procedure is followed, determining baseline and gain constants by fitting Gaussian distributions on $0\unit{\e}$ and $1\unit{\e}$ peaks, and applying these to the image to convert from raw signal in ADU to charge in $\unit{\e}$.

To compute the ARSC rate, selection cuts from~\cite{sensei_1e_rate} are used to remove backgrounds from high-energy events and local Skipper-CCD defects.
The pixels passing the cuts are used to compute the mean charge density $\mu$ from all images under each data taking condition.

\subsubsection{\label{sec:srsc_protocol}Serial register spurious charge}
After pumping in the serial register, the serial register is read out along with additional pixels in the overscan region. This process is repeated multiple times to form a serial register image, where each row consists of one array of serial register pixels.
Skipper readout with $800$ measurement samples per pixel is performed to achieve a readout noise of $\sim0.10\unit{\e}$.
Column binning is employed to reduce readout time.
Calibration of each serial register image is performed in the same way as in ARSC measurements. After calibration, the following selection cuts are then applied in order:
\begin{itemize}
    \item Overscan pixels:
    Remove non-physical pixels in the overscan region and restrict the analysis to physical pixels in the serial register.
    \item Serial register hits:
    Remove high energy particle interactions in the serial register or nearby field-free region that produce horizontal strips of high-charge pixels, termed serial register hits~\cite{serialregisterhit}.
    An entire row of serial register pixels is removed if a serial register hit is found above the baseline charge density in that row.
    \item Hot images:
    We check for serial register images that deviate from the average charge density to ensure consistent performance. We do not find any such images in these datasets.
    \item Hot pixels:
    Remove pixel locations that consistently deviate from the average charge density across all serial register images in a run.
\end{itemize}
The pixels passing the selection criteria are used to compute a charge density $\mu$, which is then fitted with Eq.~(\ref{eq:sc_fit}) to extract a spurious charge rate $R_{\text{SC}}$.
With these cuts, the data selection efficiencies for measurements with the detector setup on the surface and at MINOS are $\sim50\%$ and $\sim95\%$, respectively.

\section{\label{sec:arsc}active region spurious charge}

We first study the properties of ARSC by operating the surface setup C-module with large vertical clock swings during pumping in the active region.
To maximize the amount of ARSC, we use the surface setup and remove the RC filter on the vertical clocks.

\begin{figure*}[!t]
    \centering
    \includegraphics[width=0.49\textwidth]{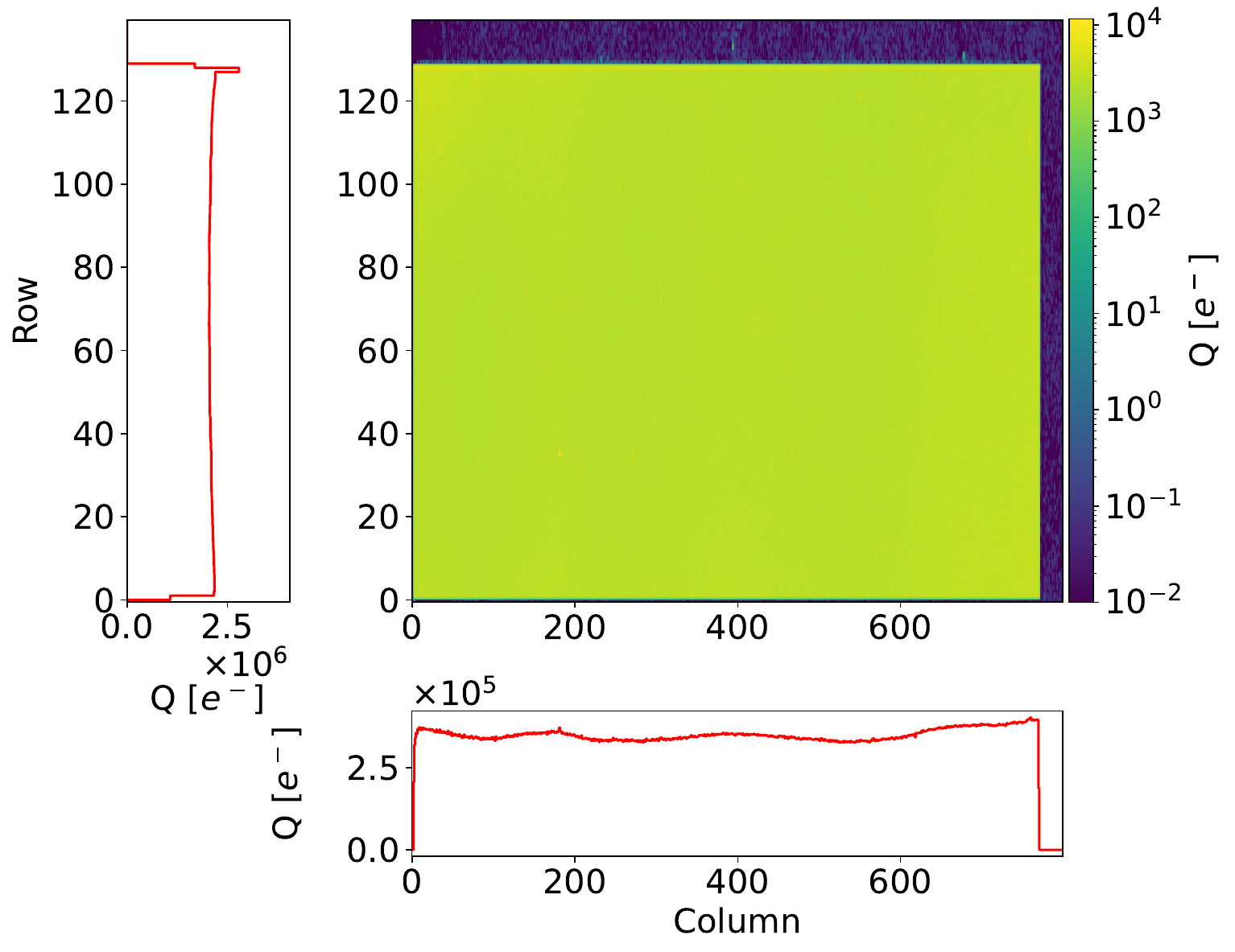}
    \hfill
    \includegraphics[width=0.49\textwidth]{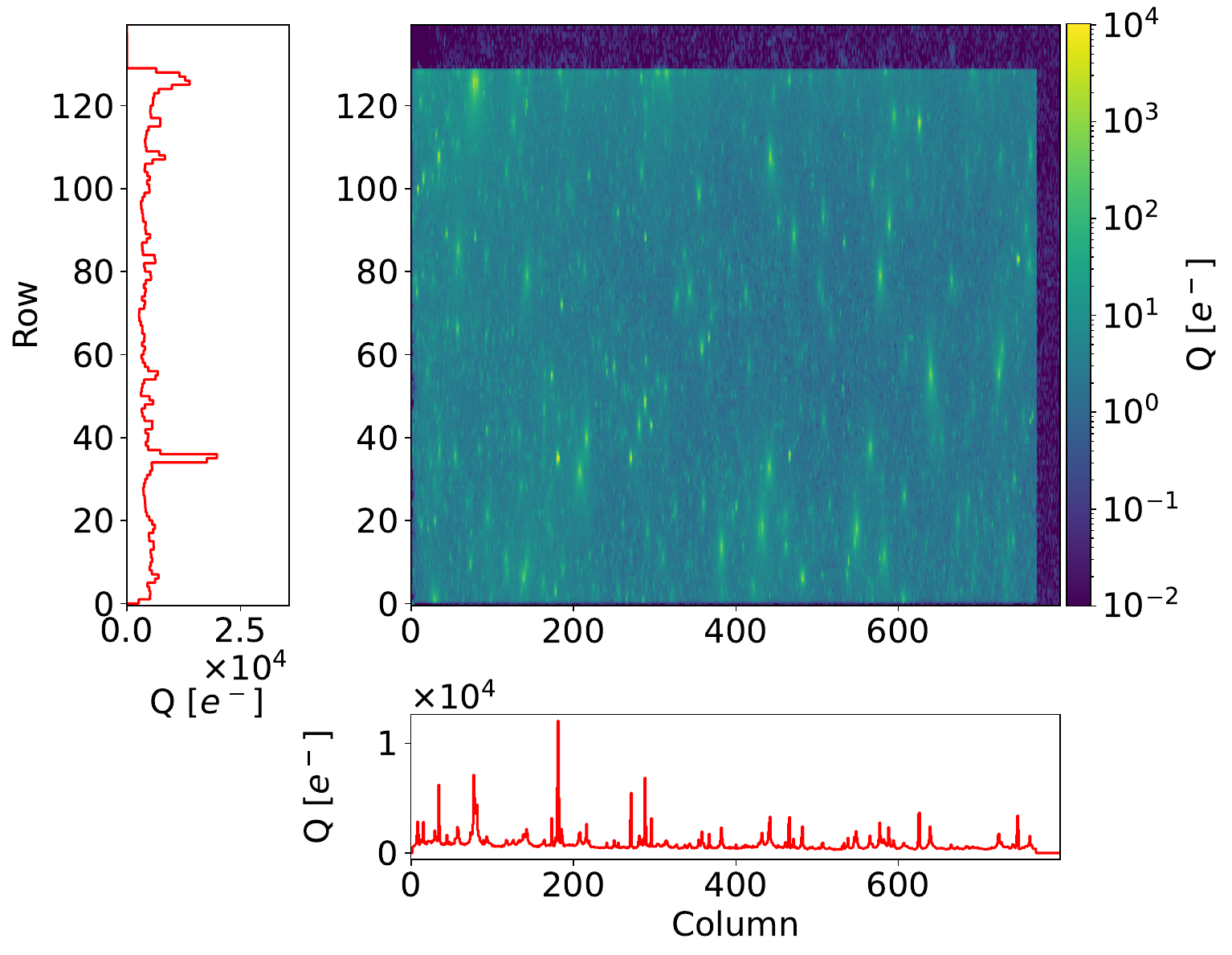}
    \caption{Calibrated images of C-module constructed by taking the median of each pixel value across a series of images to remove high-energy tracks. Each series of images is read out with $4\times4$ binning after $200000$ active region pixel-pumps with vertical clock swings of $10.25\unit{V}$. \textbf{Left}: Skipper-CCD initialized with the all-clocks purge procedure described in the text before pumping. The pixel-pumping results in a distribution of charge with approximately $3000$\unit{\e/superpixel}. \textbf{Right}: Skipper-CCD initialized with the vertical-only purge procedure before pumping. The charge density is reduced to $\mathcal{O}(10)$\unit{\e/superpixel}. Localized excess charge consistent with clocking luminescence becomes apparent with reduced ARSC. The overscan regions are also visible in both images, exhibiting negligible charge. Additional $x-$ and $y-$ projections show the summed charge per superpixel column and superpixel row.}
    \label{fig:arsc}
\end{figure*}

Fig.~\ref{fig:arsc} shows two calibrated images taken after all-clocks and vertical-only purge respectively, read with $4\times4$ binning after $200{,}000$ pixel-pumps in the active region with vertical clock swings of $10.25\unit{V}$.
The pumping and readout take about $100\unit{s}$ and $27\unit{minutes}$, respectively. The amount of charge generated by dark current is small compared to the ARSC.
In the left panel, a charge of approximately 3000\unit{\e/superpixel} is observed due to active region pixel-pumping after an all-clocks purge initialization.
Conversely, if a vertical-only purge initialization is used prior to active region pixel-pumping, the amount of charge is reduced to $\mathcal{O}(10)$\unit{\e/superpixel}.
In tests of e-purges with several different sets of clocks, this reduction of ARSC was observed whenever the vertical clocks were lowered while the transfer gate was kept at its nominal operating voltage.

To explain this drastic reduction in the observed charge when applying the vertical-only purge procedure, we hypothesize that in a vertical-only purge, electrons in the parallel channel stops are pushed from the active region to the transfer gate by the potential difference, and they then drain to the backside of the Skipper-CCD.
A parallel channel stop and the channels of the two adjacent pixel columns form a parasitic static induction transistor with the parallel channel stop as static induction transistor source, neighboring channels as gate, and the backside layer as drain~\cite{steve_2006,steve_2009}.
The turn-on threshold of the static induction transistor depends on the width of the parallel channel stop. Because this width is greater where it overlaps the transfer gate ($5\unit{\mu m}$) than in the active area ($3\unit{\mu m}$), the static induction transistor can more easily drain channel stop electrons from the transfer gate. Consequently, pushing electrons to the transfer gate removes them from the channel stop and from ARSC generation.

The charge generated by active-region pumping after vertical-only purge (see right panel of Fig.~\ref{fig:arsc}) appears in a fixed pattern of diffuse ``dots.'' The dots are also present after all-clocks purge, but less prominent.
The width of the dots is larger than expected from charge diffusion in the bulk of the Skipper-CCD and does not depend on the substrate voltage.
The amount of charge increases with the delay time between each pixel transfer, as was previously observed in \cite{emccd}.
All of these observations are consistent with the standard explanation of what is known as ``clocking luminescence''~\cite{janesick2001scientific} or ``Jim Dots''~\cite{jimdots}, in which charges at local defects are energized during clocking and emit infrared light that, in turn, is absorbed in nearby pixels.

In addition to the pumping method discussed above done with the surface setup C-module, we also measure ARSC under typical SENSEI Skipper-CCD readout conditions using the SENSEI@MINOS detector with skinny modules.
We perform these measurements with the pixel-pumping technique and data selection described in Section~\ref{sec:meas}, utilizing the standard SENSEI RC filters and vertical clock swings of $2.25\unit{V}$, and initializing with all-clocks purge.
A delay time of $\sim47\unit{s}$, corresponding to the time to read out a row of a C-module image with Skipper readout, is added between each vertical transfer. 
Under this condition, the ARSC is measured at $(2.3\pm1.1)\times10^{-8}\unit{\e/pixel/transfer}$.
Assuming that the ARSC per transfer is consistent between skinny and C-modules, this extrapolates to
a single-electron density of $\sim6\times10^{-6}\unit{\e/pixel/image}$ for a C-module, which is small compared to the overall exposure-independent single-electron density.
In addition to this measurement performed with all-clocks purge, we perform a measurement with vertical-only purge. However, unlike our observation reported above with the surface C-module and a $10.25\unit{V}$ vertical clock swing, where the vertical-only purge led to a drastic reduction in measured ARSC, we here do not find any reduction when applying the vertical-only purge and when using the standard SENSEI vertical clock swing of $2.25\unit{V}$.

\section{\label{sec:srsc} Serial register spurious charge}

In this section, we investigate the spurious charge generated in the serial register.  In \S\ref{subsec:sourcesSRSC}, we discuss several possible sources for the SRSC.  In \S\ref{subsec:delay},  we investigate when in the readout process the spurious charge is generated.  
Then, motivated by these measurements, we discuss and demonstrate the tri-level clocking scheme in \S\ref{subsec:mitigation}, which mitigates the SRSC.

\subsection{Sources of serial register spurious charge}\label{subsec:sourcesSRSC}

Both parallel channel stops and the serial register channel stop are present near the serial register (see Fig.~\ref{fig:ccd_structure}).
To determine whether SRSC originates from the parallel channel stops or the serial register channel stop, we compare the amount of SRSC generated in the prescan region (column $\le8$) with non-prescan (column $>8$) regions of the serial register, since only the non-prescan region of the serial register is adjacent to the parallel channel stops.

\begin{figure*}[!t]
    \centering
    \includegraphics[width=0.49\textwidth]{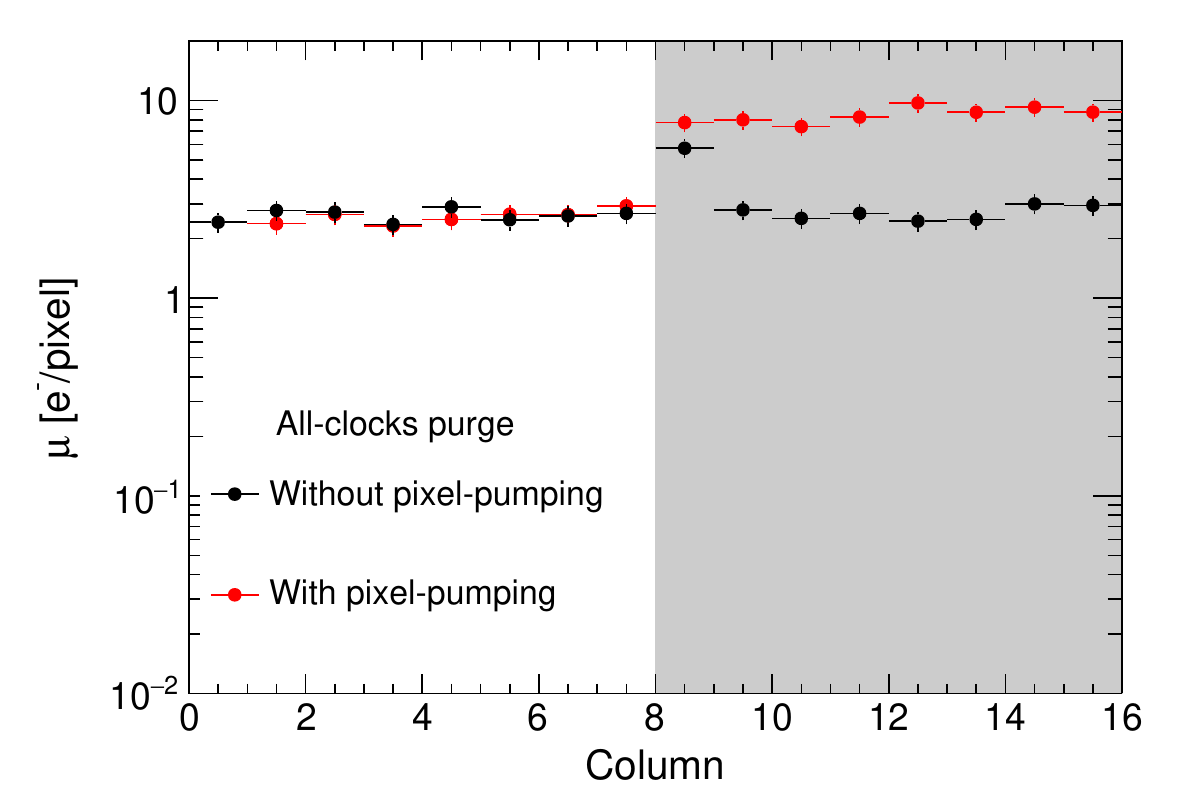}
    \hfill
    \includegraphics[width=0.49\textwidth]{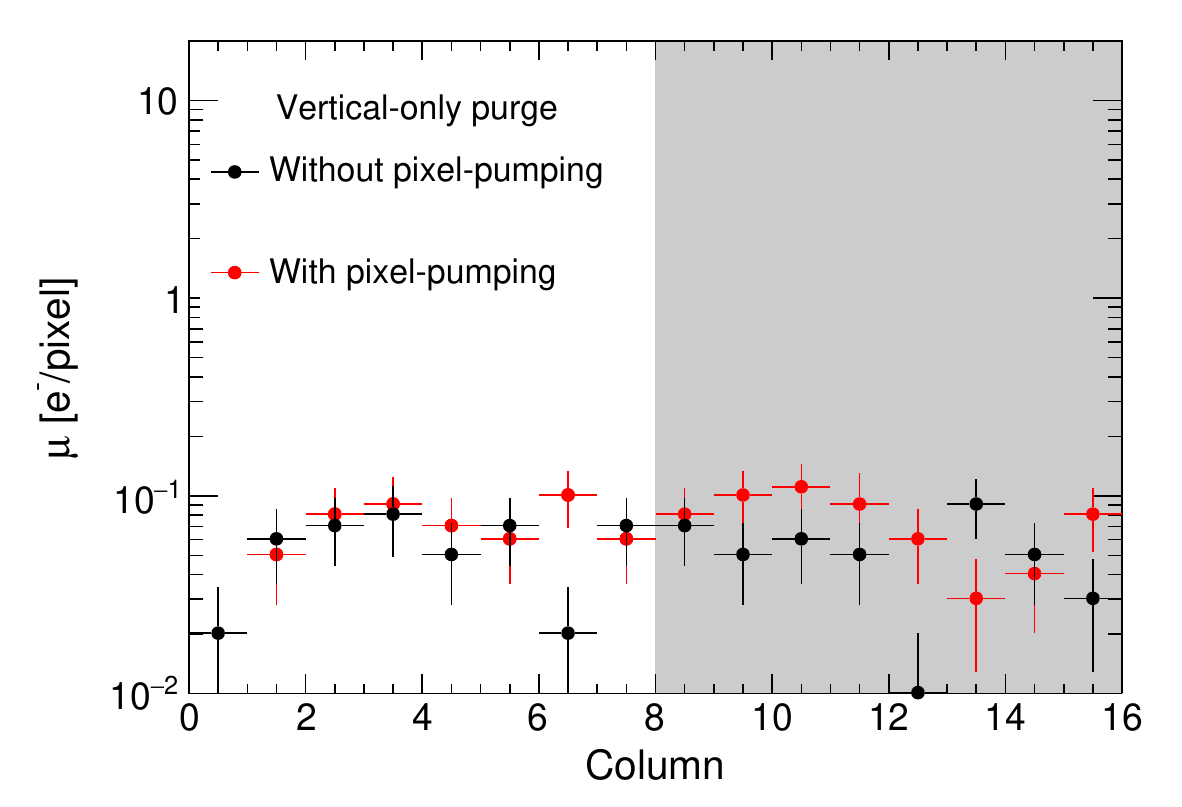}
    \caption{
    Serial register charge density distribution of a C-module in surface setup without pixel-pumping  (black) and with pixel-pumping (red) using horizontal clock swings of $6\unit{V}$ at surface setup following either the all-clocks purge procedure (\textbf{left}) or the vertical-only purge procedure (\textbf{right}). The prescan region corresponds to  column numbers $\leq8$, while the non-prescan region has column numbers $>8$ and is shaded gray. 
    With the pixel-pumping procedure, the first transfer is in the direction of the readout stage; consequently, the first pixel in the serial register is drained, resulting in zero charge at readout. 
    Under all-clocks purge, pixel-pumping creates a visible excess only in the non-prescan region, demonstrating that the parallel channel stops produce the majority of SRSC under this condition. This excess charge does not appear under vertical-only purge, because that reduces the SRSC produced by parallel channel stops. 
    The vertical-only purge procedure also reduces the measured charge across all columns that is obtained without pumping, showing that this charge is mostly SRSC generated when charge packets traverse the non-prescan region.
    }
    \label{fig:pump_prescan}
\end{figure*}

We perform this measurement using the surface setup with the C-module. Fig.~\ref{fig:pump_prescan} shows the serial register charge density distribution with and without pixel-pumping using horizontal clock swings of $6$\unit{V}, under either the all-clocks purge (left) or vertical-only purge (right) initialization. 
In all images, SRSC is also generated during cleaning and readout; every charge packet in the serial register image is transferred through the full length of the serial register corresponding to that quadrant.
As an example, the charge packet at position $x$ underwent $3080-x$ transfers during cleaning and $x$ transfers during readout.

Without pixel-pumping, both prescan and non-prescan regions have similar charge densities as expected. Under the all-clocks purge initialization, the pixel-pumping results in an elevated amount of charges in the non-prescan region, while the prescan region charge density remains consistent with the charge density obtained when no pixel-pumping is performed. This demonstrates that under these conditions, parallel channel stops are the main source of SRSC, and the serial register channel stops do not produce significant SRSC.
In two out of four quadrants, excess charge was observed at the boundary of the prescan and non-prescan regions, both with and without pixel-pumping; this remains the subject of further study.

With vertical-only purge, pixel-pumping does not produce excess charge in either region. This confirms that vertical-only purge reduces the SRSC generated by parallel channel stops with large clock swings, similar to what is observed for ARSC.

\begin{figure}[!t]
\centering
\includegraphics[scale=0.43]{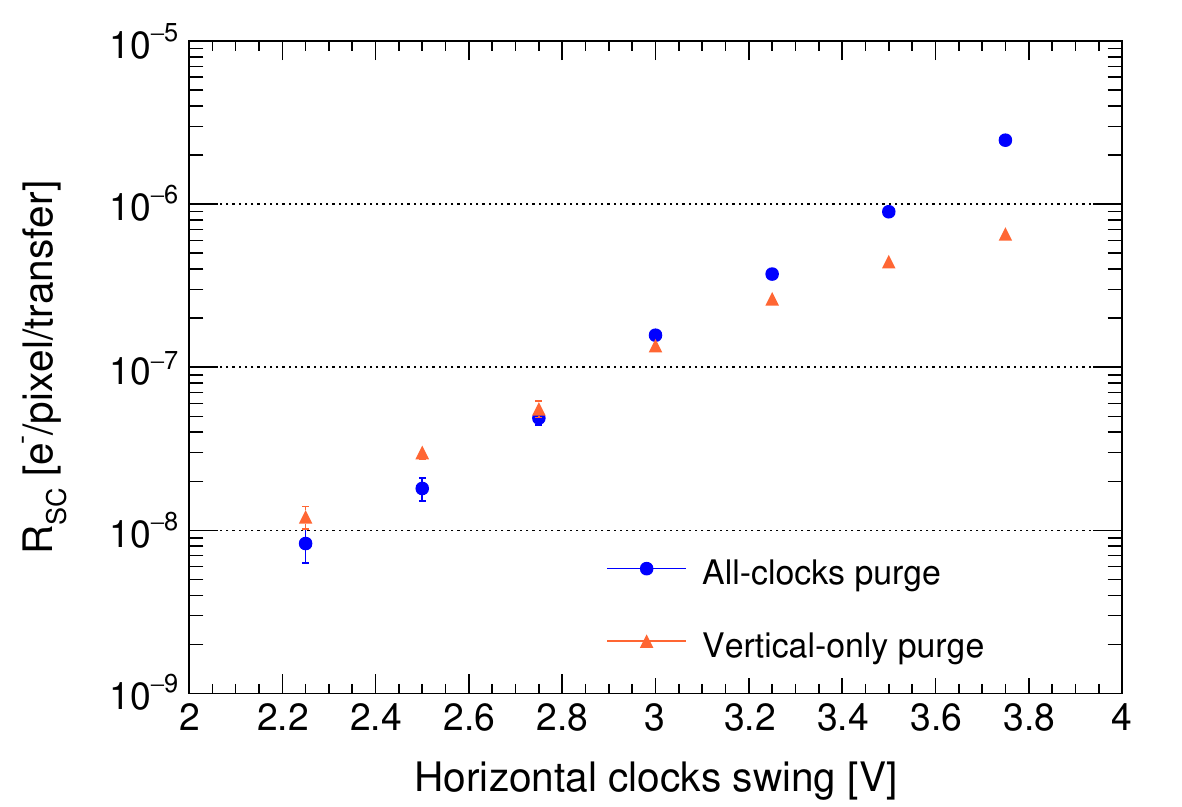}
\caption{Serial register spurious charge (SRSC) rate measured with the pixel-pumping method at various horizontal clock swings following (blue) all-clocks purge and (red) vertical-only purge. A delay of $17.8\unit{ms}$ is added between each horizontal transfer during pumping. 
An SRSC reduction from vertical-only purge is only seen at horizontal clock swings higher than $3\unit{V}$. This measurement is performed with SENSEI@MINOS detector with C-module.}
\label{fig:sc_swing_epurge}
\end{figure}

To investigate the source of SRSC under SENSEI Skipper-CCD operations, we would like to measure the effect of e-purge on the serial register at horizontal clock swings of $2.25\unit{V}$.
However, at the resulting reduced SRSC rate, the small number of pixels in the prescan region makes it difficult to repeat the comparison of prescan and non-prescan regions due to low statistics.
Instead, we use the SENSEI@MINOS detector with a C-module to measure the rate of SRSC generated in the entire serial register after all-clocks purge and vertical-only purge, with horizontal clock swings ranging from $2.25\unit{V}$ to $3.75\unit{V}$ (see Fig.~\ref{fig:sc_swing_epurge}). 
A delay time of $17.8\unit{ms}$ is added between each horizontal transfer during pumping. This delay time corresponds to the typical Skipper readout time of a single pixel.
The vertical-only purge significantly reduces the SRSC compared to all-clocks purge at horizontal clock swings greater than $3\unit{V}$, which suggests that at these swings, under all-clocks purge, the majority of SRSC is coming from parallel channel stops. This is consistent with the conclusion from comparing the prescan and non-prescan regions at $6\unit{V}$.
This reduction is not observed at horizontal clock swings less than $3\unit{V}$; it appears that at low clock swings, neither SRSC nor ARSC is significantly reduced by the vertical-only purge.
Also, since the SRSC at low horizontal clock swings is similar when initializing with the vertical-only purge and the all-clocks purge, and since we cannot measure the prescan region due to low statistics, we cannot determine whether in this case the parallel channel stops, or the serial register channel stop, or both are the source of SRSC.

\subsection{Effect of clocking delays on spurious charge}\label{subsec:delay}

\begin{figure}[!t]
\centering
\includegraphics[width=1\textwidth]{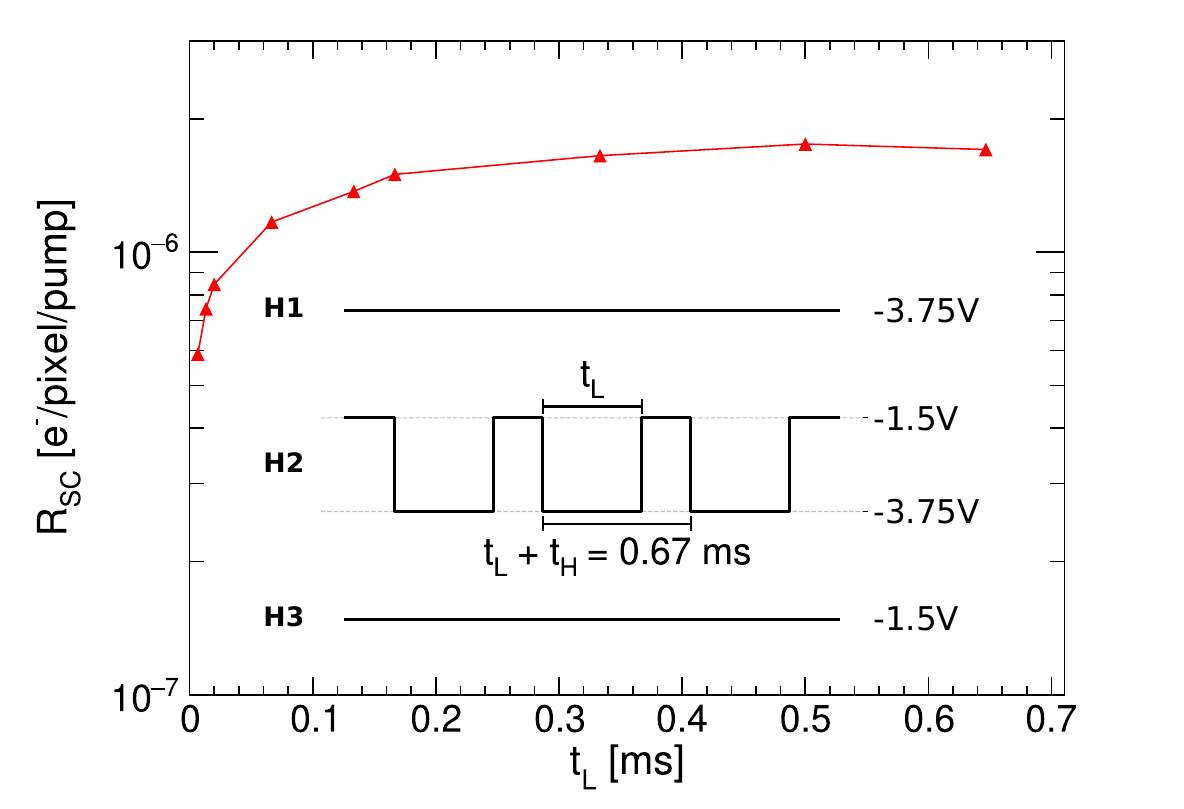}
\caption{Serial register spurious charge (SRSC) generated per phase transition as a function of low voltage pulse width $t_{\text{L}}$. The high voltage pulse width $t_{\text{H}}$ is adjusted accordingly so the total period is $0.67$\unit{ms}. The measurement was taken using the single-phase-pumping method, where one phase (H2) is pumped and, to prevent charge from moving between pixels, the other two (H1 and H3) are held low and high, respectively. The horizontal clock swing was $4\unit{V}$ and the CCD was initialized using all-clocks purge. The increase in SRSC with increasing low voltage pulse width is consistent with the model that trapped electrons are responsible for producing SRSC. This measurement is performed with the surface setup with the C-module.}
\label{fig:sc_phase_delay_time}
\end{figure}

We investigated the polarity of the trapped charges responsible for SRSC generation, as well as the dependence of SRSC generation on the delay time following each phase transition, using the single-phase-pumping method on the surface setup with the C-module.
Using horizontal clock swings of $4\unit{V}$ and all-clocks purge, and varying the low and high voltage durations while keeping a total period of $0.67\unit{ms}$, the amount of SRSC generated per phase transition is observed to increase with the horizontal clock low voltage pulse width (see Fig.~\ref{fig:sc_phase_delay_time}). 
This measurement is consistent with the model described in Section~\ref{sec:sc} that SRSC is generated by trapped electrons released when the clock voltage is low.

During the normal data-taking operation of SENSEI Skipper-CCDs, the delay for each horizontal clock transition during a serial register pixel transfer is $20\unit{\mu s}$, and the total time to transfer the charge of a single serial register pixel is $120\unit{\mu s}$.
In comparison, the Skipper readout of one pixel takes around $14\unit{ms}$ to $18\unit{ms}$ depending on the setup and the desired readout noise for good separation between $0$\e and $1$\e peaks ($\sim0.14$\unit{\e}).
Since the horizontal clocks are kept constant while the readout stage reads the newly transferred charge packet, this delay between horizontal pixel transfers is several orders of magnitude longer than the horizontal pixel transfer time itself and has a significant impact on the rate of SRSC generated in Skipper-CCDs.

\begin{figure}[!t]
\centering
\includegraphics[scale=0.43]{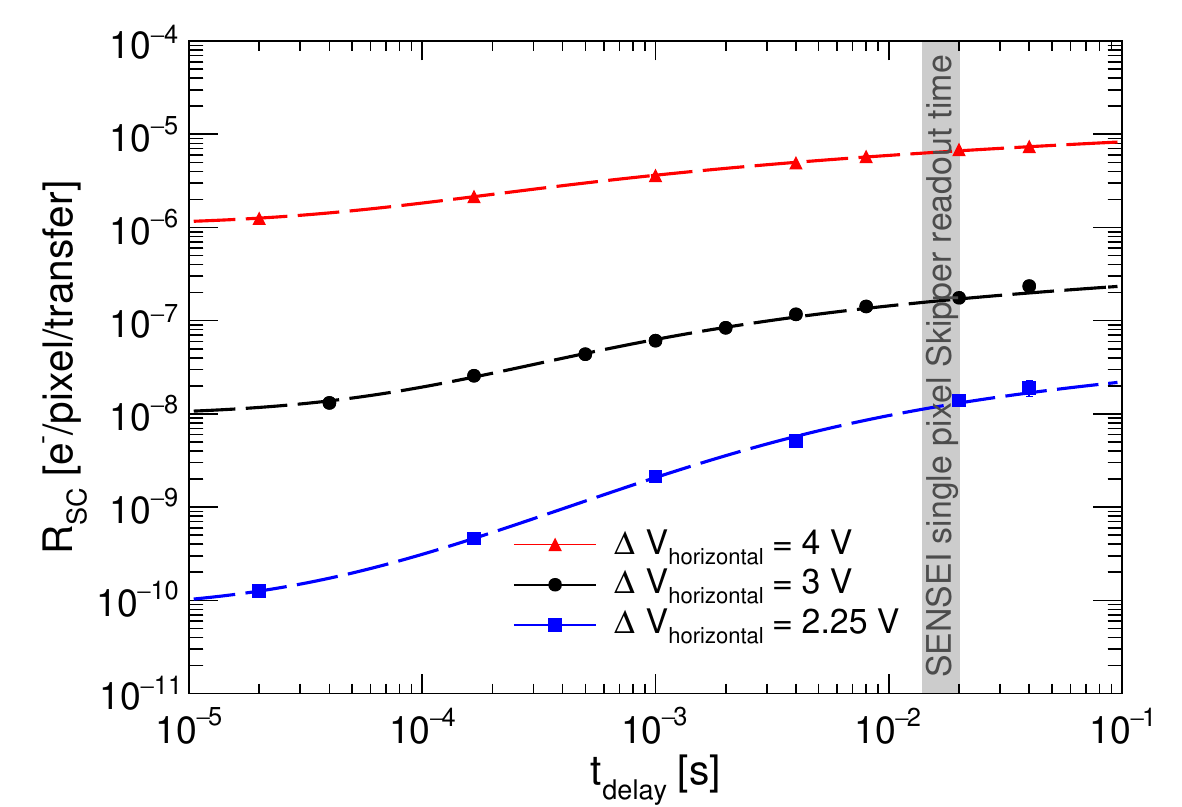}
\caption{Serial register spurious charge generated per pixel transfer, as a function of the delay time $t_{\text{delay}}$ between each pixel transfer for three different horizontal clock swings. 
The measurement is performed with SENSEI@MINOS detector with C-module.
The experimental data are fitted with the logarithmic function in Eq.~(\ref{eq:sc_log}).
The fitted values of $\tau_{\text{min}}$ are $248\pm61$\unit{\mu s}, $35\pm3$\unit{\mu s}, and $8.8\pm0.4$\unit{\mu s} for $2.25$\unit{V}, $3$\unit{V}, and $4$\unit{V} horizontal clock swings, respectively.
The typical SENSEI readout time of a single pixel is shaded in gray.}
\label{fig:sc_delay_fit}
\end{figure}

To characterize the impact of delay time between pixel transfer on how much SRSC is generated, we measure the SRSC rate using the pixel-pumping method with a variable delay time, $t_{\text{delay}}$, added after each pixel transfer.
The results from the SENSEI@MINOS detector with a C-module, for three values of the horizontal clock swing, are shown in Fig.~\ref{fig:sc_delay_fit}. 
For each horizontal clock swing, the SRSC rate as a function of $t_{\text{delay}}$ is well described by a logarithmic function 
\begin{equation}
 R_{\text{SC}} = A\cdot \ln\left(1+\frac{t_{\text{delay}}}{\tau_{\text{min}}}\right) + C \,,
 \label{eq:sc_log}
\end{equation}
which corresponds to Eq.~(\ref{eq:sc_model}) under the assumption that the trap emission time distribution $f(\tau_{\text{e}})$ has a lower bound at $\tau_{\text{min}}$, but then behaves as $1/\tau_{\text{e}}$.
The constant $A$ and $C$ characterize the SRSC dependence on horizontal clock swings and the SRSC generated during pixel transfer, respectively.
We observe that $\tau_{\text{min}}$ decreases as the horizontal clock swing increases, suggesting that the trap emission times are shorter.
This suggests that at higher horizontal clock swings, the trapped electrons that contribute to SRSC experience a higher electric field, and their emission time constant is reduced by field-enhanced emission effects, such as Poole-Frenkel~\cite{Electric_field_effect_on_thermal_emission,Electric_field_enhanced_emission,charge_trap_poole-frenkel}.
 
The SRSC dependence on the delay time impacts the effectiveness of RC filtering, which is a standard method of reducing SC~\cite{janesick2001scientific}, particularly at high horizontal clock swings~\cite{Villalpando_2024}.
Shaping the clock waveform with an RC filter reduces the peak electric field and thus the amount of SRSC that is generated at times comparable to the RC filter time constant, but the RC time constant must be kept short compared to the pixel transfer time to allow the clock to reach its target voltage.
With Skipper-CCDs, we can therefore suppress spurious charge generated up to the pixel transfer time but cannot suppress the spurious charge generated between that time and the pixel readout time.
With higher horizontal clock swings, where more of the SRSC is generated at shorter times, RC filtering is relatively effective.
However, with low horizontal clock swings, very little of the SRSC is generated on short time scales where RC filtering is effective.
As shown in Fig.~\ref{fig:sc_delay_fit}, at $\Delta V_{\text{H}}=2.25\unit{V}$, only $\sim 1$\% of the SRSC accumulated during a full Skipper readout is produced during the pixel transfer itself ($t_{\text{delay}}=0$\unit{s}). The remaining $\sim 99$\% is generated during the dwell time while the Skipper amplifier samples the pixel. 

Next-generation Skipper-CCDs with multiple in-line amplifiers or SiSeRO amplifiers are able to achieve sub-electron readout noise with shorter readout time~\cite{mas_ccd,sisero}; we therefore expect less SRSC generated during Skipper readout. For the current generation of Skipper-CCDs, SRSC can be reduced by either reducing horizontal clock swings or shaping the horizontal clock waveforms during Skipper readout.

\subsection{Mitigating spurious charge generation in the serial register with a tri-level clocking scheme}\label{subsec:mitigation}

The horizontal clocks swing of $2.25\unit{V}$ used in SENSEI has already been optimized for low SRSC; a further reduction in clock swing during pixel transfer results in significant charge transfer inefficiency (CTI). 
However, we have shown that under these conditions SRSC is mostly produced during Skipper readout, in the horizontal phase that is held low.
Raising the voltage of that phase at that time can reduce SRSC.
This reduces the potential well depth during Skipper readout and consequently the full-well capacity of the serial register, but the full-well capacity is not a critical performance parameter in rare event searches, where signal events consist of only a few electron-hole pairs~\cite{Essig:2011nj,rouven_subgev}.

\begin{figure}[!t]
\centering
\includegraphics[width=\textwidth]{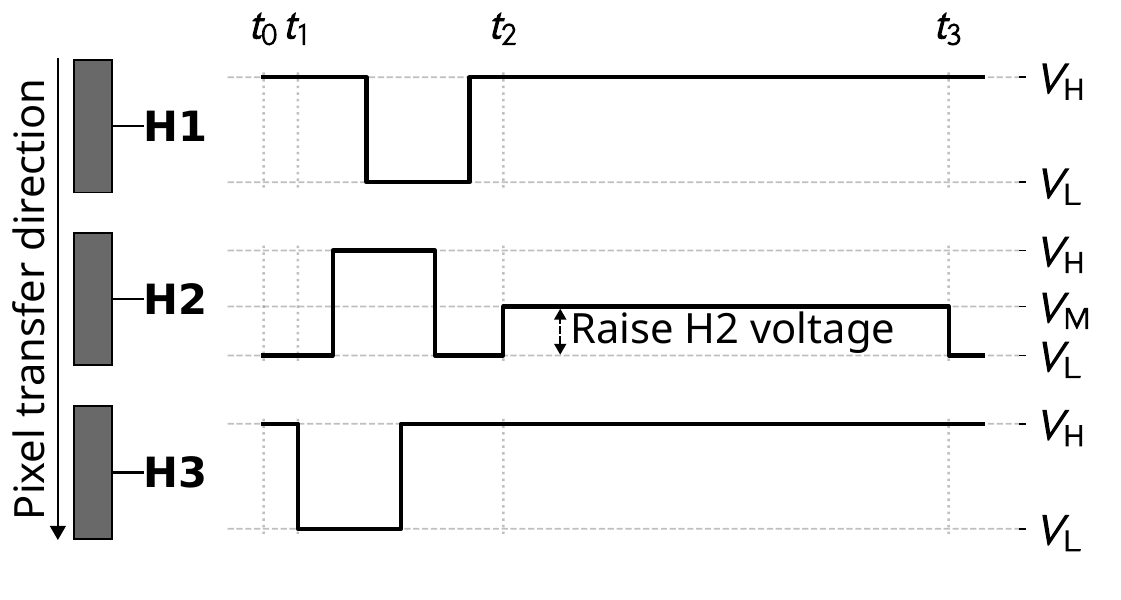}
\caption{Tri-level clocking sequence during pixel transfer and Skipper readout. At $t_0$, the charge packet is held in a potential well beneath the H2 gate of each pixel. From $t_1$ to $t_2$, the charge packet is transferred to the next pixel in the H3 direction, with full horizontal clock swings of $V_{\text{H}}-V_{\text{L}}$. During the Skipper readout from $t_2$ to $t_3$, the H2 voltage increases to a midlevel voltage $V_{\text{M}}$ to reduce the potential well. After Skipper readout is completed, H2 drops back to $V_{\text{L}}$ before the next pixel transfer.}
\label{fig:trilevel_diagram}
\end{figure}

We thus develop a serial register clocking scheme where the low phase is raised during Skipper readout (see Fig.~\ref{fig:trilevel_diagram}).
During pixel transfer, the horizontal clocks have the full swing of $V_{\text{H}}-V_{\text{L}}$, as in the normal clocking scheme. 
Once the horizontal transfer is completed and the Skipper readout starts, the H2 gate voltage would normally stay at $V_{\text{L}}$ to hold the charge packet in place.
We instead raise the H2 voltage slightly to an intermediate value, $V_{\text{M}}$.
At the end of Skipper readout, the H2 gate voltage returns to $V_{\text{L}}$ to prepare for subsequent pixel transfer.
We refer to this as ``tri-level clocking'' for Skipper-CCDs, by analogy to the tri-level clocking used to reduce spurious charge in other CCDs~\cite{janesick2001scientific}; in that scheme, the mid-level voltage is used during pixel transfers.

We implemented the tri-level clocking in SENSEI@MINOS with the C-module. 
To optimize the tri-level clocking to minimize SRSC without impacting CTI, we took several single-electron resolution images at various potential well depths during Skipper readout. The optimized potential well during Skipper readout was found to be at $V_{\text{H}}-V_{\text{M}}=1.2\unit{V}$, the smallest value where the CTI was visibly identical to that with normal clocking.

To measure the impact of tri-level clocking on the serial register single-electron density, we took serial register images under both normal clocking and tri-level clocking schemes. In these serial register images, pixels in the serial register were repeatedly read out without transferring charge from the active region.
While the pumping measurement is specifically designed to isolate SRSC, the serial register images provide a more direct measurement of the total single-electron density in the serial register, as opposed to extrapolating the pumping measurement.
We applied the same cuts described in Section~\ref{sec:srsc_protocol} on these serial register images.

\begin{figure}[!t]
\centering
\includegraphics[width=\textwidth]{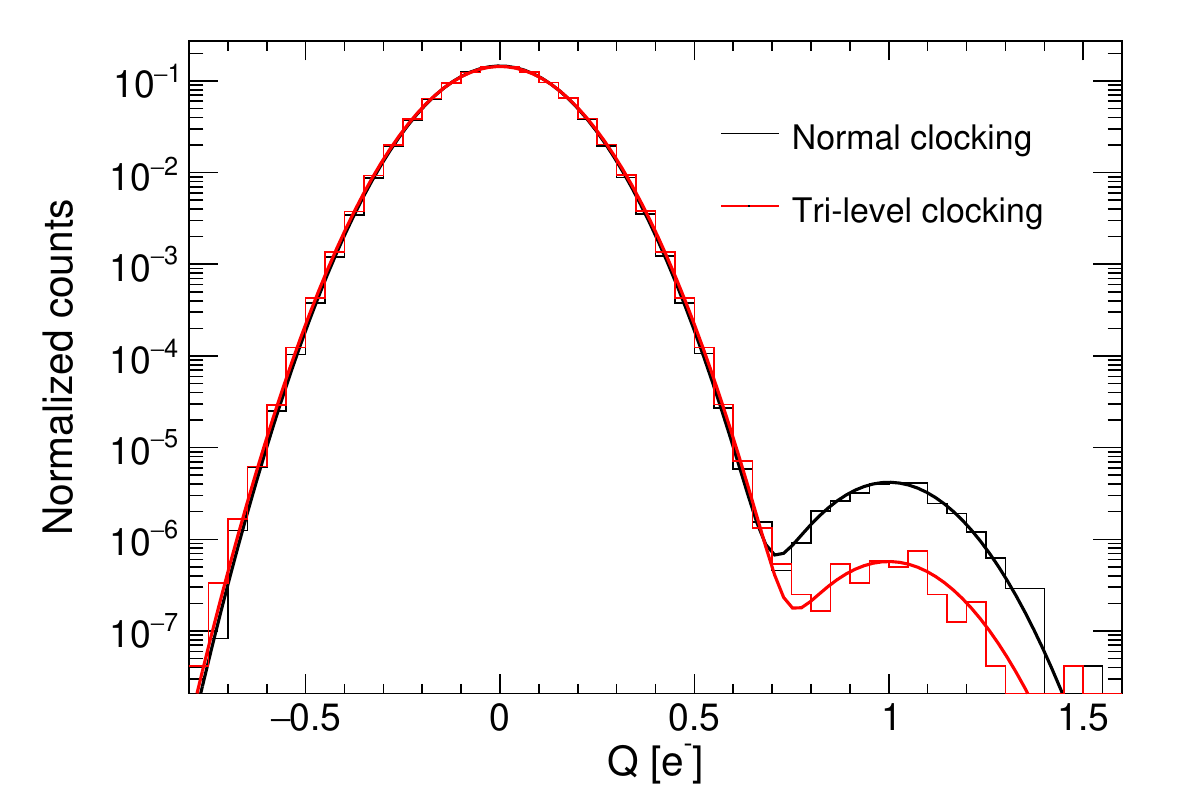}
\caption{Charge distribution of serial register pixels readout with normal clocking (black) and tri-level clocking with a $1.2\unit{V}$ potential well during Skipper readout (red) using the SENSEI@MINOS detector with the C-module. Tri-level clocking reduces the single-electron density in the serial register from $(2.9\pm0.1)\times 10^{-5}\unit{\e/pixel/image}$ to $(4.0\pm0.4)\times 10^{-6}\unit{\e/pixel/image}$.}
\label{fig:sc_trilevel_result}
\end{figure}

The charge distribution of serial register pixels with both normal clocking and tri-level clocking of the C-module at SENSEI@MINOS with all quadrants combined are shown in Fig.~\ref{fig:sc_trilevel_result}.
With normal clocking, the measured single-electron density in the serial register is $(2.9\pm0.1)\times 10^{-5}\unit{\e/pixel/image}$.
We note that extrapolating the SRSC measurement obtained from pixel-pumping ($\sim1\times10^{-8}\unit{\e/pixel/transfer}$ from Fig.~\ref{fig:sc_swing_epurge}) yields a single-electron density of $\sim1.5\times10^{-5}\unit{\e/pixel/image}$.
For tri-level clocking with a potential well depth of $1.2\unit{V}$, we observe a serial register single-electron density of $(4.0\pm0.4)\times 10^{-6}\unit{\e/pixel/image}$.
This demonstrates that tri-level clocking is a promising clocking technique for reducing SRSC in Skipper-CCDs for low background experiments.

\section{\label{sec:impact}Summary}
In this work, we characterize spurious charge generation in both the active region and the serial register. We observed that the majority of spurious charge in Skipper-CCDs is produced during Skipper readout when vertical and horizontal gate voltages are held constant.
We find that spurious charge at high clock swings is reduced if the parallel channel stops are purged by lowering the voltage on the vertical clocks while keeping the transfer gate at its nominal operating voltages, though this has negligible benefit at the low clock swings typically used in rare event searches.
A tri-level clocking scheme, where the low phase is raised during Skipper readout, has shown significant improvement in reducing single-electron backgrounds in the serial register, offering a promising technique to improve sensitivity of Skipper-CCDs in future rare event searches.

Future efforts will focus on quantifying the contributions from the spurious charge from the active region and the serial register to the overall exposure-independent single-electron density in the readout conditions used for rare-event searches.
To further determine the source of serial register spurious charge, future measurements with multi-amplifier-sensing Skipper-CCDs~\cite{mas_ccd} may enable additional measurements of spurious charge in the prescan region at low horizontal clock swings.

\begin{acknowledgments}
We thank James Janesick for useful discussions on spurious charge and clocking luminescence.
We are grateful for the support of the Heising-Simons Foundation (HSF) under Grant No.~79921. 
This document was prepared by the SENSEI collaboration using the resources of the Fermi National Accelerator Laboratory (Fermilab), a U.S. Department of Energy, Office of Science, Office of High Energy Physics HEP User Facility. Fermilab is managed by Fermi Forward Discovery Group, LLC, acting under Contract No. 89243024CSC000002.
The CCD development work was supported in part by the Director, Office of Science, of the DOE under No.~DE-AC02-05CH11231. AD and YW acknowledge support from the Universities Research Association and HSF Grant No.~79921. 
RE and YW acknowledge support from the Simons Investigator in Physics award MPS-SIP-00010469. RE also acknowledges support from DOE Grant DE-SC0025309. ADW acknowledges support from NASA APRA award No.\ 80NSSC22K1411 and a grant from the Heising-Simons Foundation (\#2023-4611).  TV acknowledge support, in part, from the Binational Science Foundation (grant No.
2024160), the Israel Science
Foundation (grant No. 2803/25) and the PAZY Foundation (grant No. 429).
Work performed at the Center for Nanoscale Materials, a U.S. Department of Energy Office of Science User Facility, was supported by the U.S. DOE, Office of Basic Energy Sciences, under Contract No. DE-AC02-06CH11357.
\end{acknowledgments}

\bibliographystyle{apsrev4-2}

\bibliography{sensei}

@article{sensei_1e_rate,
  title = {SENSEI at SNOLAB: Single-Electron Event Rate and Implications for Dark Matter},
  author = {Bloch, Itay M. and Botti, Ana M. and Cababie, Mariano and Cancelo, Gustavo and Cervantes-Vergara, Brenda A. and Daal, Miguel and Desai, Ansh and Drlica-Wagner, Alex and Essig, Rouven and Estrada, Juan and Etzion, Erez and Moroni, Guillermo Fernandez and Holland, Stephen E. and Kehat, Jonathan and Lawson, Ian and Luoma, Steffon and Orly, Aviv and Perez, Santiago E. and Rodrigues, Dario and Saffold, Nathan A. and Scorza, Silvia and Sofo-Haro, Miguel and Stifter, Kelly and Tiffenberg, Javier and Uemura, Sho and Villalpando, Edgar Marrufo and Volansky, Tomer and Winkel, Federico and Wu, Yikai and Yu, Tien-Tien},
  collaboration = {SENSEI Collaboration},
  journal = {Phys. Rev. Lett.},
  volume = {134},
  issue = {16},
  pages = {161002},
  numpages = {7},
  year = {2025},
  month = {Apr},
  publisher = {American Physical Society},
  doi = {10.1103/PhysRevLett.134.161002},
  url = {https://link.aps.org/doi/10.1103/PhysRevLett.134.161002}
}

@article{damic_probing,
  title = {Probing Benchmark Models of Hidden-Sector Dark Matter with DAMIC-M},
  author = {Aggarwal, K. and Arnquist, I. and Avalos, N. and Bertou, X. and Castell\'o-Mor, N. and Chavarria, A. E. and Cuevas-Zepeda, J. and Dastgheibi-Fard, A. and De Dominicis, C. and Deligny, O. and Duarte-Campderros, J. and Estrada, E. and Ga\"{\i}or, R. and Gkougkousis, E.-L. and Hossbach, T. and Iddir, L. and Kavanagh, B. J. and Kilminster, B. and Lantero-Barreda, A. and Lawson, I. and Letessier-Selvon, A. and Lin, H. and Loaiza, P. and Lopez-Virto, A. and Lou, R. and McGuire, K. J. and Munagavalasa, S. and Noonan, J. and Norcini, D. and Paul, S. and Privitera, P. and Robmann, P. and Roach, B. and Settimo, M. and Smida, R. and Traina, M. and Vilar, R. and Yajur, R. and Venegas-Vargas, D. and Zhu, C. and Zhu, Y.},
  collaboration = {DAMIC-M Collaboration},
  journal = {Phys. Rev. Lett.},
  volume = {135},
  issue = {7},
  pages = {071002},
  numpages = {10},
  year = {2025},
  month = {Aug},
  publisher = {American Physical Society},
  doi = {10.1103/2tcc-bqck},
  url = {https://link.aps.org/doi/10.1103/2tcc-bqck}
}

@article{sensei_snolab,
  title = {First Direct-Detection Results on Sub-GeV Dark Matter Using the SENSEI Detector at SNOLAB},
  author = {Adari, Prakruth and Bloch, Itay M. and Botti, Ana M. and Cababie, Mariano and Cancelo, Gustavo and Cervantes-Vergara, Brenda A. and Crisler, Michael and Daal, Miguel and Desai, Ansh and Drlica-Wagner, Alex and Essig, Rouven and Estrada, Juan and Etzion, Erez and Moroni, Guillermo Fernandez and Holland, Stephen E. and Kehat, Jonathan and Korn, Yaron and Lawson, Ian and Luoma, Steffon and Orly, Aviv and Perez, Santiago E. and Rodrigues, Dario and Saffold, Nathan A. and Scorza, Silvia and Singal, Aman and Sofo-Haro, Miguel and Stefanazzi, Leandro and Stifter, Kelly and Tiffenberg, Javier and Uemura, Sho and Villalpando, Edgar Marrufo and Volansky, Tomer and Wu, Yikai and Yu, Tien-Tien and Emken, Tim\'on and Xu, Hailin},
  collaboration = {SENSEI Collaboration},
  journal = {Phys. Rev. Lett.},
  volume = {134},
  issue = {1},
  pages = {011804},
  numpages = {8},
  year = {2025},
  month = {Jan},
  publisher = {American Physical Society},
  doi = {10.1103/PhysRevLett.134.011804},
  url = {https://link.aps.org/doi/10.1103/PhysRevLett.134.011804}
}

@article{sensei_2020,
  title = {SENSEI: Direct-Detection Results on sub-GeV Dark Matter from a New Skipper CCD},
  author = {Barak, Liron and Bloch, Itay M. and Cababie, Mariano and Cancelo, Gustavo and Chaplinsky, Luke and Chierchie, Fernando and Crisler, Michael and Drlica-Wagner, Alex and Essig, Rouven and Estrada, Juan and Etzion, Erez and Moroni, Guillermo Fernandez and Gift, Daniel and Munagavalasa, Sravan and Orly, Aviv and Rodrigues, Dario and Singal, Aman and Haro, Miguel Sofo and Stefanazzi, Leandro and Tiffenberg, Javier and Uemura, Sho and Volansky, Tomer and Yu, Tien-Tien},
  collaboration = {SENSEI Collaboration},
  journal = {Phys. Rev. Lett.},
  volume = {125},
  issue = {17},
  pages = {171802},
  numpages = {7},
  year = {2020},
  month = {Oct},
  publisher = {American Physical Society},
  doi = {10.1103/PhysRevLett.125.171802},
  url = {https://link.aps.org/doi/10.1103/PhysRevLett.125.171802}
}

@book{janesick2001scientific,
  title={Scientific Charge-Coupled Devices},
  author={Janesick, James R},
  volume={83},
  year={2001},
  publisher={SPIE Press}
}

@misc{lta,
      title={Low Threshold Acquisition controller for Skipper CCDs}, 
      author={Gustavo Cancelo and Claudio Chavez and Fernando Chierchie and Juan Estrada and Guillermo Fernandez Moroni and Eduardo Emilio Paolini and Miguel Sofo Haro and Angel Soto and Leandro Stefanazzi and Javier Tiffenberg and Ken Treptow and Neal Wilcer and Ted Zmuda},
      year={2020},
      eprint={2004.07599},
      archivePrefix={arXiv},
      primaryClass={astro-ph.IM},
      url={https://arxiv.org/abs/2004.07599}, 
}

@article{skipper1e,
  title = {Single-Electron and Single-Photon Sensitivity with a Silicon Skipper CCD},
  author = {Tiffenberg, Javier and Sofo-Haro, Miguel and Drlica-Wagner, Alex and Essig, Rouven and Guardincerri, Yann and Holland, Steve and Volansky, Tomer and Yu, Tien-Tien},
  journal = {Phys. Rev. Lett.},
  volume = {119},
  issue = {13},
  pages = {131802},
  numpages = {6},
  year = {2017},
  month = {Sep},
  publisher = {American Physical Society},
  doi = {10.1103/PhysRevLett.119.131802},
  url = {https://link.aps.org/doi/10.1103/PhysRevLett.119.131802}
}

@ARTICLE{steve_2003,
  author={Holland, S.E. and Groom, D.E. and Palaio, N.P. and Stover, R.J. and Mingzhi Wei},
  journal={IEEE Transactions on Electron Devices}, 
  title={Fully depleted, back-illuminated charge-coupled devices fabricated on high-resistivity silicon}, 
  year={2003},
  volume={50},
  number={1},
  pages={225-238},
  doi={10.1109/TED.2002.806476}
  }

@ARTICLE{steve_2009,
  author={Holland, Stephen E. and Kolbe, William F. and Bebek, Christopher J.},
  journal={IEEE Transactions on Electron Devices}, 
  title={Device Design for a 12.3-Megapixel, Fully Depleted, Back-Illuminated, High-Voltage Compatible Charge-Coupled Device}, 
  year={2009},
  volume={56},
  number={11},
  pages={2612-2622},
  doi={10.1109/TED.2009.2030631}
  }

@inproceedings{steve_2006,
author = {S. E. Holland and C. J. Bebek and K. S. Dawson and J. H. Emes and M. H. Fabricius and J. A. Fairfield and D. E. Groom and A. Karcher and W. F. Kolbe and N. P. Palaio and N. A. Roe and G. Wang},
title = {{High-voltage-compatable fully depleted CCDs}},
volume = {6276},
booktitle = {High Energy, Optical, and Infrared Detectors for Astronomy II},
editor = {David A. Dorn and Andrew D. Holland},
organization = {International Society for Optics and Photonics},
publisher = {SPIE},
pages = {62760B},
year = {2006},
doi = {10.1117/12.672393},
URL = {https://doi.org/10.1117/12.672393}
}

@article{Villalpando_2024,
doi = {10.1088/1538-3873/ad2865},
url = {https://doi.org/10.1088/1538-3873/ad2865},
year = {2024},
month = {apr},
publisher = {The Astronomical Society of the Pacific},
volume = {136},
number = {4},
pages = {045001},
author = {Villalpando, Edgar Marrufo and Drlica-Wagner, Alex and Plazas Malagón, Andrés A. and Bakshi, Abhishek and Bonati, Marco and Campa, Julia and Cancino, Braulio and Chavez, Claudio R. and Estrada, Juan and Fernandez Moroni, Guillermo and Fraga, Luciano and Gaido, Manuel E. and Holland, Stephen and Hur, Rachel and Jonas, Michelle and Moore, Peter and Tiffenberg, Javier},
title = {Characterization and Optimization of Skipper CCDs for the SOAR Integral Field Spectrograph},
journal = {Publications of the Astronomical Society of the Pacific}
}

@article{burkegajar,
  title = {Dynamic Suppression of Interface-State Dark Current in Buried-Channel {{CCDs}}},
  author = {Burke, B.E. and Gajar, S.A.},
  month = {2},
  year={1991},
  journal = {IEEE Transactions on Electron Devices},
  volume = {38},
  number = {2},
  pages = {285--290},
  issn = {1557-9646},
  doi = {10.1109/16.69907},
  url = {https://ieeexplore.ieee.org/document/69907},
  eventtitle = {{{IEEE Transactions}} on {{Electron Devices}}},
}

@article{e2v_purge,
   title={Tearing and related field distortions in deep-depletion charge-coupled devices},
   volume={5},
   ISSN={2329-4124},
   url={http://dx.doi.org/10.1117/1.JATIS.5.4.041505},
   DOI={10.1117/1.jatis.5.4.041505},
   number={04},
   journal={Journal of Astronomical Telescopes, Instruments, and Systems},
   publisher={SPIE-Intl Soc Optical Eng},
   author={Juramy, Claire and Antilogus, Pierre and Le Guillou, Laurent and Sepulveda, Eduardo},
   year={2019},
   month={9},
   pages={1} }

@article{sensei_mariano,
  title = {SENSEI: Characterization of Single-Electron Events Using a Skipper Charge-Coupled Device},
  author = {Barak, Liron and Bloch, Itay M. and Botti, Ana and Cababie, Mariano and Cancelo, Gustavo and Chaplinsky, Luke and Chierchie, Fernando and Crisler, Michael and Drlica-Wagner, Alex and Essig, Rouven and Estrada, Juan and Etzion, Erez and Fernandez Moroni, Guillermo and Gift, Daniel and Holland, Stephen E. and Munagavalasa, Sravan and Orly, Aviv and Rodrigues, Dario and Singal, Aman and Haro, Miguel Sofo and Stefanazzi, Leandro and Tiffenberg, Javier and Uemura, Sho and Volansky, Tomer and Yu, Tien-Tien},
  collaboration = {SENSEI Collaboration},
  journal = {Phys. Rev. Appl.},
  volume = {17},
  issue = {1},
  pages = {014022},
  numpages = {9},
  year = {2022},
  month = {Jan},
  publisher = {American Physical Society},
  doi = {10.1103/PhysRevApplied.17.014022},
  url = {https://link.aps.org/doi/10.1103/PhysRevApplied.17.014022}
}

@article{sensei_modulation,
  title = {SENSEI: A search for diurnal modulation in sub-GeV dark matter scattering},
  author = {Bloch, Itay M. and Botti, Ana M. and Cababie, Mariano and Cancelo, Gustavo and Cervantes-Vergara, Brenda A. and Daal, Miguel and Desai, Ansh and Drlica-Wagner, Alex and Essig, Rouven and Estrada, Juan and Etzion, Erez and Moroni, Guillermo Fernandez and Holland, Stephen E. and Kehat, Jonathan and Lawson, Ian and Luoma, Steffon and Orly, Aviv and Perez, Santiago E. and Rodrigues, Dario and Saffold, Nathan A. and Scorza, Silvia and Sofo-Haro, Miguel and Stifter, Kelly and Tiffenberg, Javier and Uemura, Sho and Villalpando, Edgar Marrufo and Volansky, Tomer and Winkel, Federico and Wu, Yikai and Yu, Tien-Tien and Bertou, Xavier},
  journal = {Phys. Rev. Lett.},
  pages = {},
  year = {2026},
  month = {Apr},
  publisher = {American Physical Society},
  doi = {10.1103/1gpp-3tqd},
  url = {https://link.aps.org/doi/10.1103/1gpp-3tqd}
}

@article{rouven_subgev,
	title = {Direct detection of sub-{GeV} dark matter with semiconductor targets},
	volume = {2016},
	issn = {1029-8479},
	url = {https://doi.org/10.1007/JHEP05(2016)046},
	doi = {10.1007/JHEP05(2016)046},
	pages = {46},
	number = {5},
	journal = {Journal of High Energy Physics},
	shortjournal = {J. High Energ. Phys.},
	author = {Essig, Rouven and Fernández-Serra, Marivi and Mardon, Jeremy and Soto, Adrián and Volansky, Tomer and Yu, Tien-Tien},
	urldate = {2022-02-05},
	year = {2016},
    month = {9},
	langid = {english}
}

@manual{LBNL_CCD_Manual,
  author       = {Bebek, C. and Roe, N.},
  title        = {4k x 2k and 4k x 4k CCD Users Manual},
  organization = {Lawrence Berkeley National Laboratory},
  year         = {2011},
  note         = {Rev. 3b}
}

@article{emccd,
author = {Nathan Bush and Julian Heymes and David Hall and Andrew Holland and Douglas Jordan},
title = {{Measurement and optimization of clock-induced charge in electron multiplying charge-coupled devices}},
volume = {7},
journal = {Journal of Astronomical Telescopes, Instruments, and Systems},
number = {1},
publisher = {SPIE},
pages = {016002},
year = {2021},
doi = {10.1117/1.JATIS.7.1.016002},
URL = {https://doi.org/10.1117/1.JATIS.7.1.016002}
}

@article{jimdots,
  title = {A {{CCD Anti-Blooming Technique}} for {{Use}} in {{Photometry}}},
  author = {Neely, A. William and Janesick, James R.},
  year = {1993},
  month = {11},
  journal = {Publications of the Astronomical Society of the Pacific},
  volume = {105},
  pages = {1330},
  publisher = {IOP},
  issn = {0004-6280},
  doi = {10.1086/133314},
  url = {https://ui.adsabs.harvard.edu/abs/1993PASP..105.1330N},
  urldate = {2026-05-27}
}

@article{sisero,
  title = {Achieving Single-Electron Sensitivity at Enhanced Speed in Fully Depleted CCDs with Double-Gate MOSFETs},
  author = {Sofo-Haro, Miguel and Donlon, Kevan and Estrada, Juan and Holland, Steve and Fahim, Farah and Leitz, Chris},
  journal = {Phys. Rev. Lett.},
  volume = {133},
  issue = {12},
  pages = {121003},
  numpages = {7},
  year = {2024},
  month = {Sep},
  publisher = {American Physical Society},
  doi = {10.1103/PhysRevLett.133.121003},
  url = {https://link.aps.org/doi/10.1103/PhysRevLett.133.121003}
}

@article{Electric_field_enhanced_emission,
    author = {Martin, P. A. and Streetman, B. G. and Hess, K.},
    title = {Electric field enhanced emission from non‐Coulombic traps in semiconductors},
    journal = {Journal of Applied Physics},
    volume = {52},
    number = {12},
    pages = {7409-7415},
    year = {1981},
    month = {12},
    issn = {0021-8979},
    doi = {10.1063/1.328731},
    url = {https://doi.org/10.1063/1.328731},
}

@article{Electric_field_effect_on_thermal_emission,
    author = {Vincent, G. and Chantre, A. and Bois, D.},
    title = {Electric field effect on the thermal emission of traps in semiconductor junctions},
    journal = {Journal of Applied Physics},
    volume = {50},
    number = {8},
    pages = {5484-5487},
    year = {1979},
    month = {08},
    issn = {0021-8979},
    doi = {10.1063/1.326601},
    url = {https://doi.org/10.1063/1.326601},
}

@article{lee_ccd,
  author   = {Du, Peizhi and Ega{\~n}a-Ugrinovic, Daniel and Essig, Rouven and Sholapurkar, Mukul},
  title    = {Low-energy radiative backgrounds in {CCD}-based dark-matter detectors},
  journal  = {Journal of High Energy Physics},
  year     = {2024},
  volume   = {2024},
  number   = {1},
  pages    = {164},
  doi      = {10.1007/JHEP01(2024)164},
  url      = {https://doi.org/10.1007/JHEP01(2024)164}
}

@article{SRH,
  title = {Statistics of the Recombinations of Holes and Electrons},
  author = {Shockley, W. and Read, W. T.},
  journal = {Phys. Rev.},
  volume = {87},
  issue = {5},
  pages = {835--842},
  numpages = {0},
  year = {1952},
  month = {Sep},
  publisher = {American Physical Society},
  doi = {10.1103/PhysRev.87.835},
  url = {https://link.aps.org/doi/10.1103/PhysRev.87.835}
}

@article{serialregisterhit,
  title = {Skipper Charge-Coupled Device for Low-Energy-Threshold Particle Experiments above Ground},
  author = {Moroni, Guillermo Fernandez and Chierchie, Fernando and Tiffenberg, Javier and Botti, Ana and Cababie, Mariano and Cancelo, Gustavo and Depaoli, Eliana L. and Estrada, Juan and Holland, Stephen E. and Rodrigues, Dario and Sidelnik, Iv\'an and Haro, Miguel Sofo and Stefanazzi, Leandro and Uemura, Sho},
  journal = {Phys. Rev. Appl.},
  volume = {17},
  issue = {4},
  pages = {044050},
  numpages = {8},
  year = {2022},
  month = {Apr},
  publisher = {American Physical Society},
  doi = {10.1103/PhysRevApplied.17.044050},
  url = {https://link.aps.org/doi/10.1103/PhysRevApplied.17.044050}
}

@ARTICLE{mas_ccd,
  author={Botti, Ana M. and Cervantes-Vergara, Brenda A. and Chavez, Claudio R. and Chierchie, Fernando and Drlica-Wagner, Alex and Estrada, Juan and Moroni, Guillermo Fernandez and Holland, Stephen E. and Irigoyen Gimenez, Blas Junior and Lapi, Agustin J. and Villalpando, Edgar Marrufo and Haro, Miguel Sofo and Tiffenberg, Javier and Uemura, Sho},
  journal={IEEE Transactions on Electron Devices}, 
  title={Single-Quantum Measurement With a Multiple-Amplifier Sensing Charge-Coupled Device}, 
  year={2024},
  volume={71},
  number={6},
  pages={3732-3738},
  doi={10.1109/TED.2024.3392711}}

@ARTICLE{charge_trap_poole-frenkel,
  author={Hall, David J. and Murray, Neil J. and Holland, Andrew D. and Gow, Jason and Clarke, Andrew and Burt, David},
  journal={IEEE Transactions on Nuclear Science}, 
  title={Determination of In Situ Trap Properties in CCDs Using a “Single-Trap Pumping” Technique}, 
  year={2014},
  volume={61},
  number={4},
  pages={1826-1833},
  doi={10.1109/TNS.2013.2295941}}

@inproceedings{ccd_pinning,
author = {James Janesick and Tom Elliott and George Frasehetti and Stewart Collins and Morley Blouke and Brian Corrie},
title = {{Charge-Coupled Device Pinning Technologies}},
volume = {1071},
booktitle = {Optical Sensors and Electronic Photography},
editor = {Morley M. Blouke and Donald Pophal},
organization = {International Society for Optics and Photonics},
publisher = {SPIE},
pages = {153 -- 169},
year = {1989},
doi = {10.1117/12.952516},
URL = {https://doi.org/10.1117/12.952516}
}

@misc{damic_new_modulation,
      title={Daily Modulation Constraints on Light Dark Matter with DAMIC-M}, 
      author={K. Aggarwal and I. Arnquist and N. Avalos and X. Bertou and N. Castello-Mor and C. Centeno-Lorca and A. E. Chavarria and J. Cuevas-Zepeda and A. Dastgheibi-Fard and C. De Dominicis and O. Deligny and J. Duarte-Campderros and E. Estrada and R. Gaior and E. -L. Gkougkousis and T. Hossbach and L. Iddir and B. J. Kavanagh and B. Kilminster and I. Lawson and A. Letessier-Selvon and H. Lin and P. Loaiza and A. Lopez-Virto and R. Lou and H. Lumengo-Kidimbu and K. J. McGuire and S. Munagavalasa and J. Noonan and 6 D. Norcini and S. Paul and P. Privitera and P. Robmann and B. Roach and D. Rosenmerkel and M. Settimo and R. Smida and M. Traina and R. Vilar and R. Yajur and D. Venegas-Vargas and C. Zhu and Y. Zhu},
      year={2025},
      eprint={2511.13962},
      archivePrefix={arXiv},
      primaryClass={hep-ex},
      url={https://arxiv.org/abs/2511.13962}, 
}

@article{damic_modu_2023,
  title = {Search for Daily Modulation of MeV Dark Matter Signals with DAMIC-M},
  author = {Arnquist, I. and Avalos, N. and Baxter, D. and Bertou, X. and Castell\'o-Mor, N. and Chavarria, A. E. and Cuevas-Zepeda, J. and Dastgheibi-Fard, A. and De Dominicis, C. and Deligny, O. and Duarte-Campderros, J. and Estrada, E. and Gadola, N. and Ga\"{\i}or, R. and Hossbach, T. and Iddir, L. and Kavanagh, B. J. and Kilminster, B. and Lantero-Barreda, A. and Lawson, I. and Lee, S. and Letessier-Selvon, A. and Loaiza, P. and Lopez-Virto, A. and McGuire, K. J. and Mitra, P. and Munagavalasa, S. and Norcini, D. and Paul, S. and Piers, A. and Privitera, P. and Robmann, P. and Scorza, S. and Settimo, M. and Smida, R. and Traina, M. and Vilar, R. and Warot, G. and Yajur, R. and Zopounidis, J-P.},
  collaboration = {DAMIC-M Collaboration},
  journal = {Phys. Rev. Lett.},
  volume = {132},
  issue = {10},
  pages = {101006},
  numpages = {7},
  year = {2024},
  month = {Mar},
  publisher = {American Physical Society},
  doi = {10.1103/PhysRevLett.132.101006},
  url = {https://link.aps.org/doi/10.1103/PhysRevLett.132.101006}
}

@article{damic_2024,
  title = {First Constraints from DAMIC-M on Sub-GeV Dark-Matter Particles Interacting with Electrons},
  author = {Arnquist, I. and Avalos, N. and Baxter, D. and Bertou, X. and Castell\'o-Mor, N. and Chavarria, A. E. and Cuevas-Zepeda, J. and Guti\'errez, J. Cortabitarte and Duarte-Campderros, J. and Dastgheibi-Fard, A. and Deligny, O. and De Dominicis, C. and Estrada, E. and Gadola, N. and Ga\"{\i}or, R. and Hossbach, T. and Iddir, L. and Khalil, L. and Kilminster, B. and Lantero-Barreda, A. and Lawson, I. and Lee, S. and Letessier-Selvon, A. and Loaiza, P. and Lopez-Virto, A. and Matalon, A. and Munagavalasa, S. and McGuire, K. J. and Mitra, P. and Norcini, D. and Papadopoulos, G. and Paul, S. and Piers, A. and Privitera, P. and Ramanathan, K. and Robmann, P. and Settimo, M. and Smida, R. and Thomas, R. and Traina, M. and Vila, I. and Vilar, R. and Warot, G. and Yajur, R. and Zopounidis, J-P.},
  collaboration = {DAMIC-M Collaboration},
  journal = {Phys. Rev. Lett.},
  volume = {130},
  issue = {17},
  pages = {171003},
  numpages = {7},
  year = {2023},
  month = {Apr},
  publisher = {American Physical Society},
  doi = {10.1103/PhysRevLett.130.171003},
  url = {https://link.aps.org/doi/10.1103/PhysRevLett.130.171003}
}

@article{Du:2020ldo,
    author = "Du, Peizhi and Egana-Ugrinovic, Daniel and Essig, Rouven and Sholapurkar, Mukul",
    title = "{Sources of Low-Energy Events in Low-Threshold Dark-Matter and Neutrino Detectors}",
    eprint = "2011.13939",
    archivePrefix = "arXiv",
    primaryClass = "hep-ph",
    doi = "10.1103/PhysRevX.12.011009",
    journal = "Phys. Rev. X",
    volume = "12",
    number = "1",
    pages = "011009",
    year = "2022"
}

@article{Essig:2011nj,
    author = "Essig, Rouven and Mardon, Jeremy and Volansky, Tomer",
    title = "{Direct Detection of Sub-GeV Dark Matter}",
    eprint = "1108.5383",
    archivePrefix = "arXiv",
    primaryClass = "hep-ph",
    reportNumber = "SLAC-PUB-14538",
    doi = "10.1103/PhysRevD.85.076007",
    journal = "Phys. Rev. D",
    volume = "85",
    pages = "076007",
    year = "2012"
}

@unpublished{CONNIE_2024,
    author = "Aguilar-Arevalo, Alexis A. and others",
    collaboration = "CONNIE",
    title = "{Searches for CE{\ensuremath{\nu}}NS and Physics beyond the Standard Model using Skipper-CCDs at CONNIE}",
    eprint = "2403.15976",
    archivePrefix = "arXiv",
    primaryClass = "hep-ex",
    reportNumber = "FERMILAB-PUB-24-0714-PPD",
    month = "3",
    year = "2024"
}

\end{document}